\documentclass{aa}

\usepackage{graphicx}
\usepackage{txfonts}
\usepackage{xcolor}
\usepackage[breaklinks,colorlinks,citecolor=blue]{hyperref}
\usepackage{soul}
\usepackage{subcaption,caption}
\usepackage{bm}
\usepackage{braket}

\usepackage{natbib}
\bibpunct{(}{)}{;}{a}{}{,}

\newcommand\be{\begin{equation}}
\newcommand\ee{\end{equation}}
\newcommand\bea{\begin{eqnarray}}
\newcommand\eea{\end{eqnarray}}
\def\ba#1\ea{\begin{align}#1\end{align}}

\newcommand\hn{\hat{n}}
\newcommand\mr{\mathrm}

\newcommand\Cl{\ensuremath{C_\ell}}

\newcommand\fsky{\ensuremath{f_\mathrm{SKY}} }

\newcommand{\Omegam}{\ensuremath{\Omega_m}}

\makeatletter
\newcommand\footnoteref[1]{\protected@xdef\@thefnmark{\ref{#1}}\@footnotemark}
\makeatother

\begin{document}

\title{Impact of survey geometry and super-sample covariance\\ on future photometric galaxy surveys}
\titlerunning{Geometry and the SSC}

\author{
S. Gouyou Beauchamps\inst{1}
\and 
F. Lacasa\inst{2}
\and 
I. Tutusaus\inst{3,4,5}
\and 
M. Aubert\inst{1,6}
\and
P. Baratta\inst{1}
\and 
A. Gorce\inst{2,7,8}
\and
Z. Sakr\inst{5, 9}
}

\institute{
Aix Marseille Univ, CNRS/IN2P3, CPPM, Marseille, France \and
Université Paris-Saclay, CNRS, Institut d'astrophysique spatiale, 91405, Orsay, France \and
Institute of Space Sciences (ICE, CSIC), Campus UAB, Carrer de Can Magrans, s/n, 08193 Barcelona, Spain \and  Institut d’Estudis Espacials de Catalunya (IEEC), Carrer Gran Capit\`a 2-4, 08034 Barcelona, Spain \and Institut de Recherche en Astrophysique et Plan\'etologie (IRAP), Universit\'e de Toulouse, CNRS, UPS, CNES, 14 Av. Edouard Belin, F-31400 Toulouse, France
\and Univ Lyon, Univ Claude Bernard Lyon 1, CNRS/IN2P3, IP2I Lyon, UMR 5822, F-69622, Villeurbanne, France \and
Department of Physics, Blackett Laboratory, Imperial College, London SW7 2AZ, U.K.\and
Department of Physics and McGill Space Institute, McGill University, Montreal, QC, Canada H3A 2T8\and 
Université St Joseph; UR EGFEM, Faculty of Sciences, Beirut, Lebanon\\
\email{gouyou@cppm.in2p3.fr}
}

\date{\today}

\abstract
{Photometric galaxy surveys probe the late-time Universe where the density field is highly non-Gaussian. A consequence is the emergence of the super-sample covariance (SSC), a non-Gaussian covariance term that is sensitive to fluctuations on scales larger than the survey window. In this work, we study the impact of the survey geometry on the SSC and, subsequently, on cosmological parameter inference. We devise a fast SSC approximation that accounts for the survey geometry and compare its performance to the common approximation of rescaling the results by the fraction of the sky covered by the survey, $f_\mr{SKY}$, dubbed "full-sky approximation". To gauge the impact of our new SSC recipe, that we call "partial-sky", we perform Fisher forecasts on the parameters of the $(w_0,w_a)$-CDM model in a 3x2 point analysis, varying the survey area, the geometry of the mask and the galaxy distribution inside our redshift bins. The differences in the marginalised forecast errors, with the full-sky approximation performing poorly for small survey areas but excellently for stage-IV-like areas, are found to be absorbed by the marginalisation on galaxy bias nuisance parameters. For large survey areas, the unmarginalised errors are underestimated by about 10\% for all probes considered. This is a hint that, even for stage-IV-like surveys, the partial-sky method introduced in this work will be necessary if tight priors are applied on these nuisance parameters. We make the partial-sky method public with a new release of the public code \texttt{PySSC}.

}

\maketitle



\section{Introduction}\label{Sect:intro}

The large-scale structure (LSS) of the Universe is an excellent probe of cosmology, giving constraints, for example, on dark matter, dark energy and the large-scale behaviour of gravity. Current galaxy surveys, such as the Kilo-Degree Survey \citep[KiDS,][]{2021A&A...646A.140H} 
and the Dark Energy Survey \citep[DES,][]{2021arXiv210513549D}, are starting to provide cosmological constraints competitive with those derived from the primordial Cosmic Microwave Background (CMB) and its weak lensing \citep{Planck2018-params}. Additionally, the next generation of stage-IV surveys, like the Vera C. Rubin Observatory Legacy Survey of Space and Time \citep[LSST,][]{Abell:2009aa}, \textit{Euclid} \citep{Euclid-redbook}, and the Dark Energy Spectroscopic Instrument \citep[DESI,][]{Aghamousa:2016zmz}, will make it possible to discriminate dark energy and modified gravity models with unprecedented precision and, doing so, their observations will shed light on the origin of cosmic acceleration.

One key challenge when studying the late-time LSS, however, is its non-Gaussian distribution, which results from the non-linear dynamics governing its time-evolution. This non-Gaussianity has a variety of consequences, at the level of the observables, as information escapes the two-point correlation function to leak into higher orders \citep[e.g.,][]{Carron2012,ObreschkowPower_2013}; at the level of the likelihood, which becomes non-Gaussian although the impact on cosmological parameters appears to be weak \citep{Lin2020,Upham2020}; and lastly at the level of the covariance, which is the main focus of this article. Non-Gaussianity typically leads to an enhanced covariance, as the tail of extreme events becomes broader. For instance, the covariance of two-point functions gets contributions from a positive trispectrum \citep[e.g.,][for galaxy clustering]{Lacasa2018b,Wadekar2019}.

One of the non-Gaussian contributions to the covariance originates in long wavelength super-sample modes, which shift the mean matter density inside the survey and modulate coherently all observables. This effect is often called sample variance in galaxy cluster analyses and super-sample covariance (SSC) when considering galaxy clustering and weak lensing. First discovered for cluster counts \citep{Hu2003}, a vast amount of literature has been devoted to the SSC in recent years \citep[e.g.,][]{Takada2013,Takada2014,Takahashi2014,Li2018,Chan2018,Lacasa2017,Lacasa2018,Barreira2018,Barreira2018b}. It is well known to have a large impact on constraints on the dark energy equation of state for future surveys, be it for clusters \citep{Hu2003}, weak lensing \citep{Barreira2018b}, or photometric galaxy clustering \cite{Lacasa2020}.

Given the fact that the SSC originates from the observation of a limited portion of the universe, the angular mask of the survey, accounting for the unobserved regions of the sky (because of bright stars or high luminosity of the Galactic plane for example), should be taken into account in the computation of the SSC. While the effect of the mask has been extensively studied for Gaussian covariance, especially for CMB analysis \citep{Hivon2002, Efstathiou2004}, there was no detailed study of whether accounting for it when including the SSC is necessary for future surveys. This is the aim of this article.

In this work, we quantify and explore the impact of the SSC on the performance of future survey missions aiming at constraining cosmology -- and, in particular, dark energy. To model the SSC, we use the approximation presented in \citet{Lacasa2019}, extend it to account for partial-sky coverage, and compare it against the often-used flat sky approximation. To forecast the impact on the inference power of the survey, we build on the \textit{Euclid} forecast efforts of \citet[hereafter EC-B2020]{Euclid2019-IST:F}. 

The article is organised as follows: in Sect.~\ref{Sect:method}, we compare three methods allowing to predict the SSC for full-sky, partial-sky, and flat-sky surveys. In Sect.~\ref{Sect:forecasts}, we describe our galaxy survey forecast methodology. In Sect.~\ref{Sect:results}, we present the impact of including the SSC on the statistical power of various survey setups in terms of signal-to-noise and Fisher forecasts. We conclude in Sect.~\ref{Sect:conclusion}.


\section{Method: SSC in full, partial and flat sky}\label{Sect:method}


Let us consider two observables $O_1$ and $O_2$. They can be written as the integral over the line-of-sight of their density, respectively $o_1$ and $o_2$: $O_i = \int \mathrm{d}V \, o_i$, where $\mr{d}V = r^{2}(z)\frac{\mr{d}r}{\mr{d}z}\mr{d}z$ is the comoving volume per steradian and $r(z)$ the comoving distance. The SSC for these observables is then given by the general formula \citep{Lacasa2019}:
\bea
\mathrm{Cov}_{\mathrm{SSC}}\left(O_{1}, O_{2}\right)= \iint \mathrm{d} V_{1} \mathrm{d} V_{2} \frac{\partial o_{1}}{\partial \delta_{b}}\left(z_{1}\right) \frac{\partial o_{2}}{\partial \delta_{b}}\left(z_{2}\right) \sigma^{2}\left(z_{1}, z_{2}\right)\,, \label{eq:def_SSC}
\eea

where the quantity $\partial o_1 / \partial \delta_b (z_1)$ describes how $o_1$ varies with changes in the background density $\delta_b$.

The (co)variance of the background density is defined as:
\bea
\nonumber \sigma^{2}(z_{1},z_{2}) &=& \langle \delta_b(z_1) \delta_b(z_2) \rangle \\
&=& \int \frac{\mathrm{d}^{3} \mathbf{k}}{(2 \pi)^{3}} \tilde{\mathcal{W}}\left(\mathbf{k}, z_{1}\right) \tilde{\mathcal{W}}^{*}\left(\mathbf{k}, z_{2}\right) P_{m}(k|z_{12})\,. \label{eq:4_sigma2_SSC_def}
\eea
where $\tilde{\mathcal{W}}$ is the Fourier transform of the survey window
function, whose expression will depend on whether we are looking at full- or partial-sky coverage. $P_{m}(k|z_{12})=D(z_1) D(z_2) P_{m}(k|z=0)$ is the matter power spectrum at redshifts $z_1$ and $z_2$.

We take $O_1$ (resp. $O_2$) to be the angular power spectrum $C^{A B}_{ij}(\ell)$ cross-correlating two LSS tracers $A$ and $B$  (resp. $C$ and $D$) -- typically, galaxy clustering and galaxy shear. Each spectra is measured for a redshift bin pair (respectively $i$ $j$ and $k$ $l$) and it can be expressed, using the Limber approximation as
\begin{equation}\label{eq:cls}
C_{ij}^{A B}\left(\ell\right)=\int \mathrm{d} V \ W_{i}^{A}(z) W_{j}^{B}(z) P_{A B}\left(k_{\ell} | z\right),
\end{equation}
where $P_{AB}(k_\ell | z)$ is the 3D power spectrum at $k_\ell \equiv (\ell+1/2)/r(z) $ and $W^A_{i}(z)$ is the kernel of observable $A$ corresponding to the redshift bin $i$. Then, from Eq.~\eqref{eq:cls} $o_{AB} = W_{i}^{A}(z) W_{j}^{B}(z) P_{A B}\left(k_{\ell} | z\right) $ (resp. $o_{CD}$) and if we assume that the derivatives $\partial o_i / \partial \delta_b$ vary slowly with redshift compared to $\sigma^2(z_1,z_2)$, we can rewrite Eq.~\eqref{eq:def_SSC} as
\begin{equation}
\begin{aligned}
\mathrm{Cov}_{\mathrm{SSC}}&( C_{ij}^{A B}(\ell), C_{kl}^{C D}(\ell) ) \simeq  \bar{\frac{\partial P_{AB}}{\partial \delta_{b}}}\left(z_{1}\right) \bar{\frac{\partial P_{CD}}{\partial \delta_{b}}} \left(z_{2}\right) \times \\ & \iint \mathrm{d} V_{1}\, \mathrm{d} V_{2} \ W_{i}^{A}(z_1) W_{j}^{B}(z_1) W_{k}^{C}(z_2) W_{l}^{D}(z_2) \, \sigma^{2}\left(z_{1}, z_{2}\right)\,.
\end{aligned}
\end{equation}
where we defined 
\begin{equation}
    \bar{\frac{\partial P_{AB}}{\partial \delta_{b}}}\left(z\right) \equiv \frac{ \int \, \mathrm{d}V\ W_{i}^{A}(z) W_{j}^{B}(z)\, \partial P_{AB} / \partial \delta_b \left(k_{\ell} | z\right) }{I^{AB}(i,j) }\,,
\end{equation}
with $I^{AB}(i,j) \equiv \int \, \mathrm{d}V\ W_{i}^{A}(z) W_{j}^{B}(z)$. Let $R^{AB}(k)$ be the effective relative response of the considered power spectrum:
\begin{equation}
    \frac{\partial P_{AB}}{\partial \delta_{b}}\left(k\right) \equiv R^{AB}(k) \, P_{AB}(k)\,.
\end{equation}
For the matter power spectrum, $R$ is constant with redshift and can be computed from perturbation theory or estimated from simulations \citep{Lacasa2019}. Then
\begin{equation}
\begin{aligned}
    \bar{\frac{\partial P_{AB}}{\partial \delta_{b}}}\left(z\right) \times I^{AB}(i,j) & = \int \, \mathrm{d}V\ W_{i}^{A}(z) W_{j}^{B}(z)\, \partial P_{AB} / \partial \delta_b \left(k_{\ell} | z\right) \\
    & = \int \, \mathrm{d}V\ W_{i}^{A}(z) W_{j}^{B}(z)\, R^{AB}(k_\ell) \, P_{AB}(k_\ell,z)\\
    & \equiv R^{AB}_\ell \,  C_{ij}^{AB}(\ell)\,.
\end{aligned}
\end{equation}

Finally, we define the matrix $S_{i, j ; k, l}^{A, B ; C, D}$, which is the dimensionless volume-averaged (co)variance of the background matter density contrast, by
\begin{equation}
\label{eq:Sijkl_matrix_def}
S_{i, j ; k, l}^{A, B ; C, D} \equiv \int \mathrm{d} V_{1} \mathrm{d} V_{2}\ \frac{W_{i}^{A}\left(z_{1}\right) W_{j}^{B}\left(z_{1}\right)}{I^{A B}\left(i, j\right)} \frac{W_{k}^{C}\left(z_{2}\right) W_{l}^{D}\left(z_{2}\right)}{I^{C D}\left(k, l\right)} \sigma^{2}\left(z_{1}, z_{2}\right)\,, 
\end{equation}
and the covariance simply rewrites as:
\begin{equation}
\label{eq:cov_exp_1}
\begin{aligned}
\mathrm{Cov}_{\mathrm{SSC}} & \left(C_{ij}^{A B}\left(\ell\right), C_{kl}^{C D}\left(\ell^{\prime}\right)\right) \approx \\ & R_{\ell}^{A B} C_{ij}^{A B}\left(\ell\right) \times R_{\ell^{\prime}}^{C D} C_{kl}^{C D}\left(\ell^{\prime}\right)  \times S_{i, j ; k, l}^{A, B ; C, D}\,.
\end{aligned}
\end{equation}

Note that here, the $\mathbf{S}$ matrix has four indices, as it describes the covariance between all auto and cross-$\Cl$'s. We denote it by $S_{ijkl}$. But for simplicityin the following we will consider only the covariance between auto-$\Cl$'s, i.e. $\mathrm{Cov}_{\mathrm{SSC}}  \left(C_{ii}^{A A}\left(\ell\right), C_{jj}^{BB}\left(\ell^{\prime}\right)\right)$. In that case the $\mathbf{S}$ matrix is denoted by $S_{ij}$. We will now derive the expression for this $\mathbf{S}$ matrix in different survey cases : full sky coverage (Sect. \ref{Sect:SSC-fullsky}), partial sky coverage (Sect. \ref{Sect:SSC-partsky}) and flat sky approximation (Sect. \ref{Sect:SSC-flatsky}).

\subsection{Full-sky case}\label{Sect:SSC-fullsky}
In the case of full-sky coverage, the variance of the background density field is simply \citep{Lacasa2016}
\begin{equation}
\begin{aligned}
\sigma^{2}(z_{1},z_{2}) & =  \dfrac{1}{2\pi^{2}} \int k^{2}\mathrm{d}k\ P_{m}(k|z_{12})\, j_{0}(kr_{1})\, j_{0}(kr_{2})\,,
\end{aligned}   
\end{equation}
where $r_i$ is the comoving distance to redshift $z_i$ and $j_0$ the spherical Bessel function of the first kind and order zero. Given that the angular matter power spectrum can be written as 
\begin{equation}
    C^{m}_{z_{1}, z_{2}}(\ell) = \dfrac{2}{\pi}\int k^{2}\mathrm{d}k\ P_{m}(k|z_{12}) \, j_{\ell}(kr_{1}) \, j_{\ell}(kr_{2})\,,
\end{equation}
we can write $\sigma^{2}$ as its monopole:
\begin{equation}
    \sigma^{2} = \dfrac{1}{4\pi}C^{m}_{z_1, z_2}(\ell=0)\,.
\end{equation}

By injecting this expression in Eq. \eqref{eq:Sijkl_matrix_def}, we can relate the $\mathbf{S}$ matrix to a power spectrum
\begin{equation}
\label{eq:4_smat_fullsky}
    S_{i, j ; k, l}^{A, B ; C, D} = \dfrac{1}{4\pi} C^{X,Y}(\ell=0)\,,
\end{equation}
where 
\ba
    a_{\ell m}^X = \dfrac{\int \mathrm{d}V\ W^{A}_{i}(z) W^{B}_{j}(z) \ a_{\ell m}^\mr{matter}(z) }{\int \mathrm{d}V\ W^{A}_{i}(z) W^{B}_{j}(z)}\,, \\
    a_{\ell m}^Y = \dfrac{\int \mathrm{d}V\ W^{C}_{k}(z) W^{D}_{l}(z) \ a_{\ell m}^\mr{matter}(z) }{\int \mathrm{d}V\ W^{C}_{k}(z) W^{D}_{l}(z)}\,,
\ea
so that
\begin{equation}
\begin{aligned}
    C^{X,Y}(\ell=0) = \int & \mathrm{d}V_{1}\mathrm{d}V_{2} \, k^{2}\mathrm{d}k \  \dfrac{W^{A}_{i}(z_{1})W^{B}_{j}(z_{1})}{I^{AB}(ij)}  \dfrac{W^{C}_{k}(z_{2})W^{D}_{l}(z_{2})}{I^{CD}(k,l)} \\
    & \times P_m(k|z_1,z_2) \ j_{0}(kr_{1}) j_{0}(kr_{2})\,.
\end{aligned}
\label{eq:angularmonopoledefinition}
\end{equation}

The exquisite sensitivity of upcoming photometric surveys will lead to a shot-noise small enough for the SSC to be an important source of error on cosmological parameters. In particular, considering a full-sky SSC such as outline above will not be sufficient anymore and the SSC associated with the limited size of the survey will need to be considered. Until now, the simple option of rescaling the full-sky covariance by a factor $f_\mr{SKY}^{-1}$, where $f_\mr{SKY} \equiv \Omega_S/4\pi$ is the fraction of the sky covered by the survey and $\Omega_S$ its solid angle, in order to account for partial-sky coverage in the SSC was used \citep{Lacasa2018}. It is the first approximation we will consider in this work.\\
Numerically, we will use the Python implementation \texttt{PySSC} \citep{Lacasa2019}\footnote{\label{note1}Available at \url{https://github.com/fabienlacasa/PySSC}.}

\subsection{Partial-sky case}\label{Sect:SSC-partsky}

In this section we derive the approach to SSC in the case of partial sky coverage.
The coverage is represented by the angular survey window function or mask $\mathcal{W}(\mathbf{\hn})$, where $\mathbf{\hn}$ is the sky direction. Typically $\mathcal{W}(\mathbf{\hn})=0$ for unobserved pixels and $\mathcal{W}(\mathbf{\hn})=1$ for observed ones, though a non-binary mask is also possible e.g. to represent the impact of inhomogeneous depth. In this article we will assume that the mask is the same at all considered redshifts. The covariance of the background mode is then given by \citep{Lacasa2018}
\be
\sigma^{2}\left(z_{1}, z_{2}\right)=\frac{1}{\Omega_{S}^{2}} \sum_{\ell}(2 \ell+1)\, C^{\mathcal{W}}(\ell)\, C^{m}_{z_{1}, z_{2}}\left(\ell \right).
\ee
As for the full-sky case, we can see the $S_{ijkl}$ matrix as a $C(\ell)$ of a non-physical field $X$, whose kernel is the product of the kernels $W^{A}W^{B}$. Here, however, multipoles other than the monopole will contribute to the SSC:
\begin{equation}
    \label{eq:4_smat_partsky}
    S^{A,B;C,D}_{i, j; k, l} = \frac{1}{\Omega_{S}^{2}} \sum_{\ell} (2\ell+1) C^{X,Y}(\ell)C^{\mathcal{W}}(\ell)\,.
\end{equation}
It is interesting to note that, when considering the full-sky limit of Eq.~\eqref{eq:4_smat_partsky} ($f_\mr{SKY} \rightarrow 1$ and $C^{\mathcal{W}}(\ell>0) = 0$), we retrieve the full-sky matrix given in Eq.~\eqref{eq:4_smat_fullsky}. 
Eq.~\eqref{eq:4_smat_partsky} is the second method for SSC estimation that we will consider in this work.

Numerically, we have developed a Python implementation of Eq.~\eqref{eq:4_smat_partsky}, building on \texttt{PySSC}. We have validated this partial-sky implementation against a parallel implementation making use of the \texttt{AngPow} \citep{Angpow:2018} public code for angular power spectra. We present this validation in Appendix \ref{Appendix_validation}, showing a $\sim 6\%$ agreement in the computation of the $S_{ij}$ for surveys larger than $\sim 2.5\%$ of the sky, while reaching $\sim 10\%$ for smaller patches.

\subsection{Flat-sky case}\label{Sect:SSC-flatsky}

Another example of approximation used in the literature to simplify the computationally-expensive estimation of the SSC is the flat-sky approximation. For a top-hat kernel of width $\delta r$ in the flat-sky case, the $S_{i,j}$ matrix can indeed be simplified to \citep{Hu2003,Lima2007}
\begin{equation}
\label{eq:flat_sky_approx}
\begin{aligned}
S_{i, j}&=\frac{1}{2 \pi^{2}} \int k_{\perp} \mathrm{d} k_{\perp} 4 \frac{J_{1}\left(k_{\perp} \theta_{S} r_{1}\right)}{k_{\perp} \theta_{S} r_{1}} \frac{J_{1}\left(k_{\perp} \theta_{S} r_{2}\right)}{k_{\perp} \theta_{S} r_{2}} \\
& \times \int \mathrm{d} k_{\|} j_{0}\left(\frac{k_{\|} \delta r_{1}}{2}\right) j_{0}\left(\frac{k_{\|} \delta r_{2}}{2}\right) \cos \left[k_{\|}\left(r_{1}-r_{2}\right)\right] P_{m}\left(k \mid z_{12}\right)\,,
\end{aligned}
\end{equation}
for a cylindrical window function of radius $\theta_S$ delineating a survey solid angle $\Omega_S = 2\pi(1-\cos \theta_S)\simeq \pi \theta_S^2$. The wave-vector $\bm{k}= (k_\parallel, k_\perp)$ is split into its components parallel and perpendicular to the line-of-sight. Here, $J_1$ is the Bessel function of the first kind and order one. The power spectrum $P_m(k \mid  z_{12})$ is evaluated at the centre of the respective redshift bins, following \citet{Hu2003,Lima2007}.\\
This approximation, very efficient computationally, will be the third and final one considered in this work. We will use it as a point of comparison only, as it is limited to the case of top-hat kernels.

\subsection{Methods comparison}\label{Sect:SSC-compare}

\begin{figure}
    \centering
    \includegraphics[width=\columnwidth]{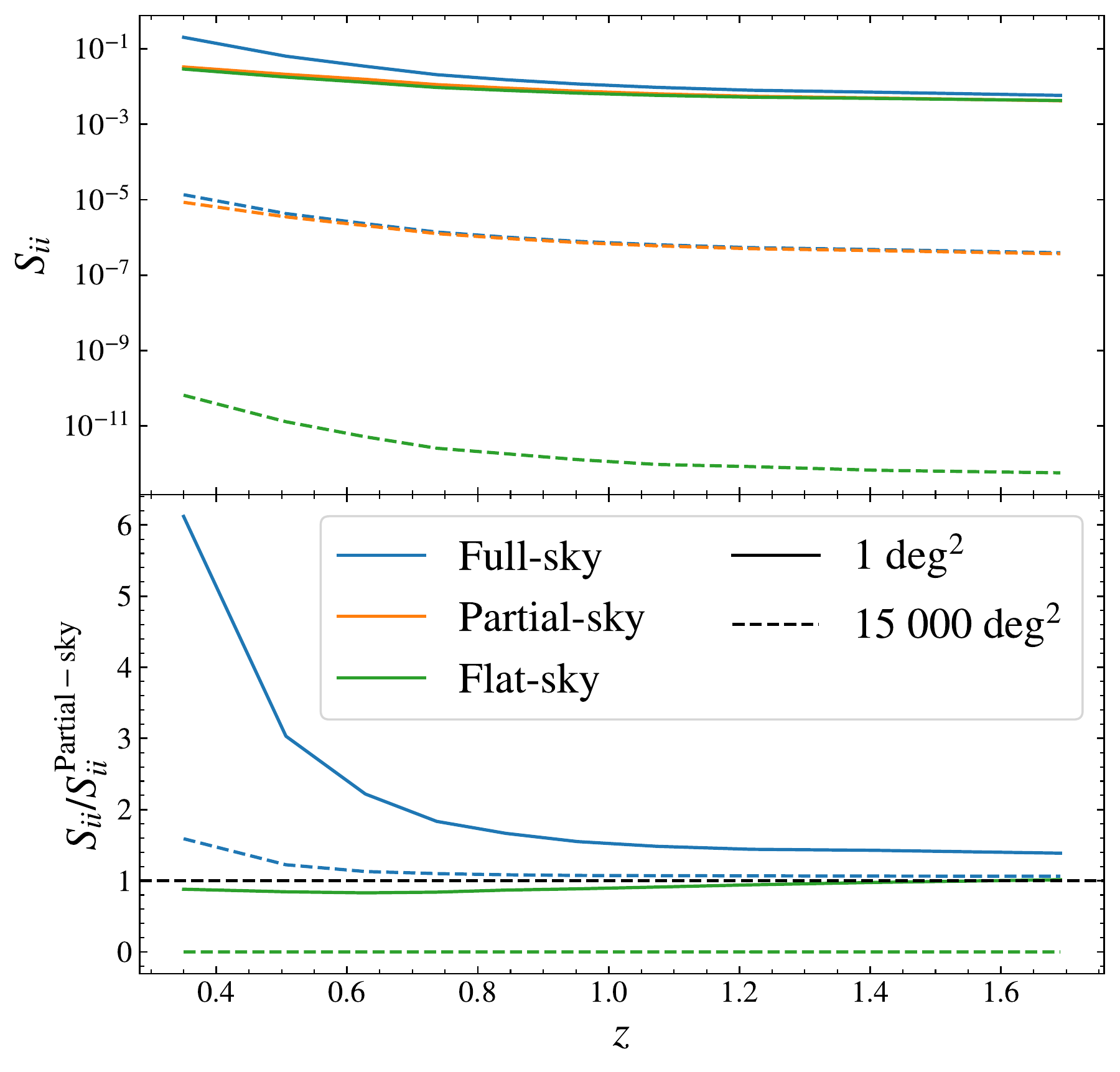}
    \caption{Diagonal terms of the $S_{ij}$ matrices obtained for surveys with area of 1 deg$^2$ (plain lines) and 15~000 deg$^2$ (dashed lines). The colours correspond to the three different methods considered in this article: full-sky in blue, partial-sky in orange and flat-sky in green. The lower panel shows the ratio to the partial-sky case.}
   \label{fig:flat_sky_comparison}
\end{figure}

In this section, we compare the $S_{ij}$ matrices obtained using the three different methods outlined in previous paragraphs, in order to choose which we will focus our interest in the rest of this work. 

\begin{figure*}
    \centering
    \includegraphics[width=.8\textwidth]{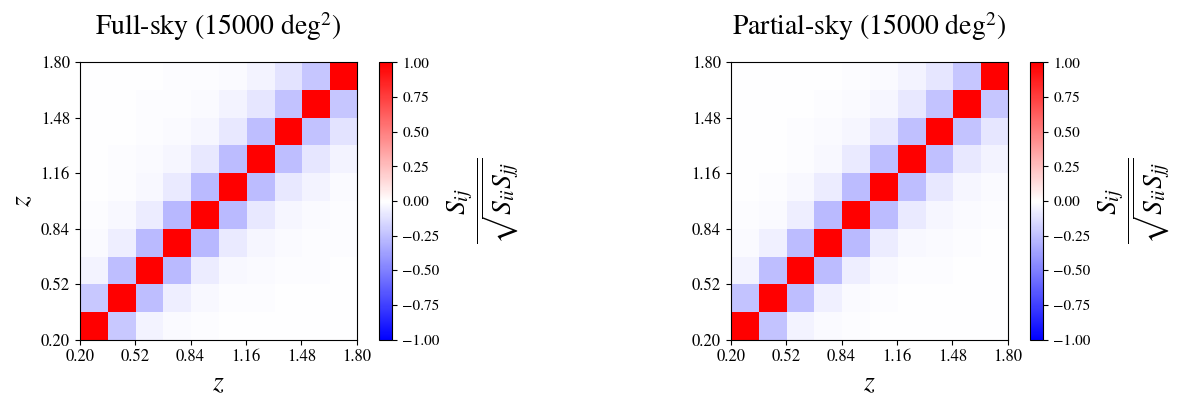}
    \caption{$S_{ij}$ correlation matrices obtained for a 15~000 square degrees circular mask and arbitrary, non overlapping, top-hat kernels for the full-sky computation of Eq.~\eqref{eq:4_smat_fullsky} (left) and the partial-sky computation of Eq.~\eqref{eq:4_smat_partsky} (right).}
    \label{fig:comp_sij_matrices}
\end{figure*}

We consider the $S_{ij}$ matrices obtained for arbitrary, top-hat kernels with the full-sky computation (Eq.~\ref{eq:4_smat_fullsky}) rescaled by $\fsky^{-1}$, the partial-sky computation (Eq.~\ref{eq:4_smat_partsky}) and with the flat-sky approximation (Eq.~\ref{eq:flat_sky_approx}), for circular masks of areas ranging from $1~\mathrm{deg}^2$ to $15~000~\mathrm{deg}^2$, as well as for 10, non overlapping, redshift bins ranging from $z=0.2$ to $z=2.0$. The choice of top-hat kernels of width $\Delta z = 0.1$ is arbitrary as this example is intended for illustration purposes only. Overall, we find on one hand that the flat-sky approximation gives a $S_{ij}$ matrix close to the true partial-sky computation for surveys with area smaller than $5~\mathrm{deg}^2$ only. On the other hand we find that the full-sky approximation, after rescaling, gives satisfying results for areas over $15~000~\mathrm{deg}^2$. 

In Fig.~\ref{fig:flat_sky_comparison}, we compare the diagonal terms of the $S_{ij}$ matrices obtained for the $1~\mathrm{deg}^2$ and $15~000~\mathrm{deg}^2$ mask for the three treatments of the mask. We find that the flat-sky approximation performs better at higher redshift, with a relative difference below $10\%$ for $z>1.5$. For lower redshifts, the $S_{ii}$ is underestimated by as much as $20\%$ when using the flat-sky approximation, which corroborates the results of \citet{Lacasa2018}. In particular, it completely ignores cross-redshift bins correlations, which however exist when using the other two approaches. For survey masks larger than $5~\mathrm{deg}^2$, the flat-sky approximation underestimates the $S_{ij}$ by as much as a factor two and by 5 orders of magnitude in the case of a wide 15~000 deg$^2$, as can be seen in Fig.~\ref{fig:flat_sky_comparison}. This approximation should, therefore, not be considered in those cases.

Conversely, the full-sky approach performs well for very wide survey masks. In this case, the full-sky approximation overestimates the diagonal of the $S_{ij}$ by up to 1.5 for low redshifts ($z<0.5$) but by less than $10\%$ for $z>1$ 

Fig.~\ref{fig:comp_sij_matrices} shows the $S_{ij}$ correlation matrices (i.e. divided by its diagonal elements) obtained for a circular survey mask of $15~000~\mathrm{deg}^2$. We can see that for non overlapping redshift bins, the SSC results in anti-correlations, which decreases for distant bins. This anti-correlation was already noted in the literature \citep{Hu2003,Lacasa2018} and comes from the fact that the matter correlation function becomes negative at large separations.

Hence, we see that both approximations recover low-redshift correlations stemming from partial-sky SSC poorly but can perform well for redshifts $z>1$, where the correlations are weaker. Here we only considered non overlapping redshift bins, but we also compared the two matrices computed for overlapping bins. In that case the correlation between the bins which overlaps becomes positive and the structure of the matrix is more complex. This renders the interpretation of the impact of the partial-sky recipe on the SSC less evident if ones only compare the $S_{ij}$ matrices. For a more complete interpretation we will see in Sect.~\ref{Sect:results} how the differences between full-sky and partial-sky directly impact cosmological parameter inference. We will not consider the flat-sky approximation further since we are interested in large cosmological surveys, with wide survey areas. Instead, we will compare results between the partial-sky computation, the full-sky approximation and the Gaussian case, in order to highlight contributions from the SSC to results.


\section{Method: galaxy surveys forecast}\label{Sect:forecasts}

In order to forecast the different constraining power of galaxy surveys depending on the covariance considered, we follow for the most part the forecast recipe presented in EC-B2020. We consider a Fisher matrix formalism and make use of the \textsc{CosmoSIS}\,\footnote{Available at \url{https://bitbucket.org/joezuntz/cosmosis/wiki/Home}.} public code\,\citep{2015A&C....12...45Z}. In this section, we review the main aspects of the forecast and refer the reader to EC-B2020
for the remaining details.

\subsection{Data}

In these forecasts, we examine the constraining power of three cosmological probes: weak lensing (WL), photometric galaxy clustering (GCph), and their cross-correlation terms (XC), also known as galaxy-galaxy lensing. We refer to the full combination with GCph + WL + XC. We consider the tomographically-binned projected angular power spectra as observables, $C_{ij}(\ell)$, where $i,j$ label redshift pairs of tomographic bins. The angular spectra have been presented in Eq.~\eqref{eq:cls} using the Limber approximation. We use the same formalism for WL, GCph, and the XC terms. The main difference between the different probes appears through the different kernels used in the projection from the power spectrum of matter perturbations, $P_{m}$ to the spherical harmonic-space observable. Following EC-B2020,  
when computing the observables, we use the Limber and flat-sky approximations \citep{Kitching2017,Kilbinger2017,Taylor2018}, and we ignore reduced shear and magnification effects \citep{Deshpande2020}.

For the redshift distribution of galaxies we follow EC-B2020 
in considering 10 tomographic redshift bins with the same number of galaxies in each bin. We assume a true underlying redshift distribution given by 
\begin{equation}\label{eq:nz1}
    n^{\rm true}(z)\propto \left(\frac{z}{z_0}\right)^2\,\text{exp}\left[-\left(\frac{z}{z_0}\right)^{3/2}\right]\,,
\end{equation}
where $z_0=0.9/\sqrt{2}$. We then compute the photometric redshift distributions in each one of the bins by convolving the true distribution with a sum of two Gaussian distributions. One for the main dispersion of photometric redshift estimates and another one for the outliers. In more detail, the redshift distribution in the tomographic bin $i$ is given by
\begin{equation}\label{eq:nz2}
    n_i(z)=\frac{\int_{z_i^-}^{z_i^+}\text{d}z_p n^{\rm true}(z)p_{ph}(z_p|z)}{\int_{z_{min}}^{z_{max}}\text{d}z\int_{z_i^-}^{z_i^+}\text{d}z_p n^{\rm true}(z)p_{ph}(z_p|z)}\,,
\end{equation}
where $(z_i^-,z_i^+)$ are the edges of the $i$th tomographic bin and set to the following values for the 10 equi-populated bins:
\begin{align}
    z_i=\{0.0010,0.42,&0.56,0.68,0.79,\nonumber\\
    &0.90,1.02,1.15,1.32,1.58,2.50\}\,.
\end{align}

We follow EC-B2020 in parameterising the probability distribution function $p_{ph}(z_p|z)$ as
\begin{align}\label{eq:nz3}
    p_{ph}(z_p|z) =&\frac{1-f_{out}}{\sqrt{2\pi}\sigma_b(1+z)}\,\text{exp}\left\{-\frac{1}{2}\left[\frac{z-c_bz_p-z_b}{\sigma_b(1+z)}\right]^2\right\}+\nonumber\\
    &+\frac{f_{out}}{\sqrt{2\pi}\sigma_o(1+z)}\,\text{exp}\left\{-\frac{1}{2}\left[\frac{z-c_oz_p-z_o}{\sigma_o(1+z)}\right]^2\right\}\,.
\end{align}

\begin{figure}[b]
\centering
\includegraphics[width=.9\linewidth]{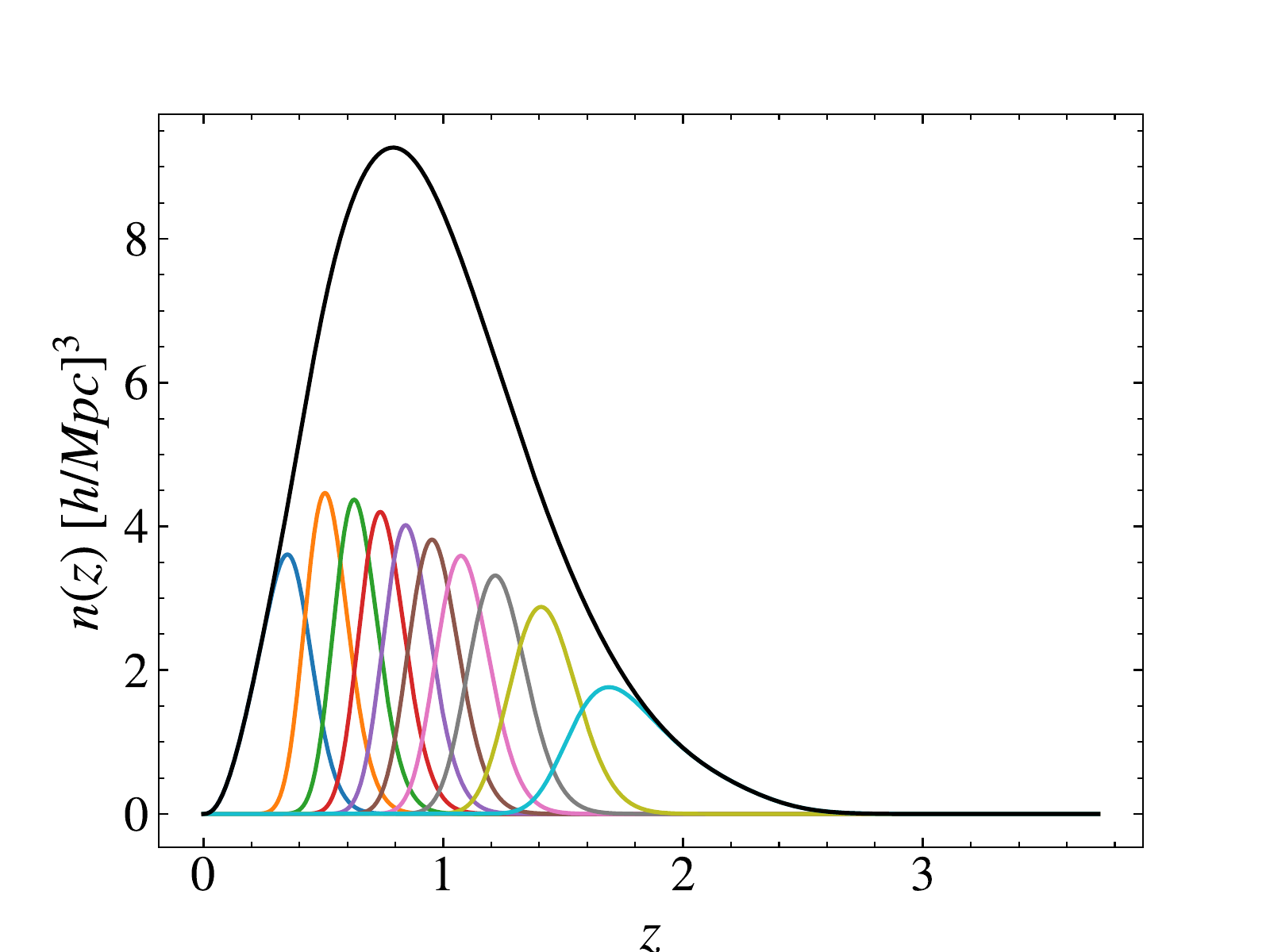}
\caption{Normalised galaxy number density distribution in the ten photometric redshift bins. The black line is the sum of all the redshift bins.}
\label{Fig:euclid_nz}
\end{figure}

For the outliers we set the multiplicative bias to $c_o=1$ and the additive bias to $z_o=0.1$. For the rest of galaxies we consider a multiplicative bias $c_b=1$ and an additive bias $z_b=0$. The uncertainty on the redshifts is assumed to be $\sigma_b=\sigma_o=0.05$. We consider a default fraction of outliers $f_{out}=0.1$. We further assume a galaxy number density of thirty galaxies per arcmin$^2$. We show in Fig.\,\ref{Fig:euclid_nz} the normalised galaxy distribution considered in this analysis.

The WL power spectra contain contributions from cosmic shear and the intrinsic alignment of galaxies. We assume these intrinsic alignments are caused by a change in galaxy ellipticities that is linear in the density field. In this case, we can express the density-intrinsic and intrinsic-intrinsic three-dimensional power spectra,  $P_{mI}$ and $P_{II}$, as a linear function of the density power spectrum, with $P_{mI} = -A(z) P_{m}$, and $P_\textrm{II} = [-A(z)]^2 P_{m}$. We follow EC-B2020 
in parameterising $A$ as
\begin{equation}
    A(z)=\frac{\mathcal{A}_{\rm IA}\mathcal{C}_{\rm IA}\Omega_m\mathcal{F}_{\rm IA}(z)}{D(z)}\,,
\end{equation}
where $\mathcal{C}_{\rm IA}=0.0134$ is a normalisation constant, $D(z)$ is the growth factor, and $\mathcal{A}_{\rm IA}$ controls the amplitude of the IA contribution. We further model the redshift dependence as
\begin{equation}
    \mathcal{F}_{\rm IA}=(1+z)^{\eta_{\rm IA}}\left[\frac{\braket{L}(z)}{L_*(z)}\right]^{\beta_{\rm IA}}\,,
\end{equation}
where $\braket{L}(z)/L_*(z)$ is the ratio between the mean source luminosity and the characteristic scale of the luminosity function \citep{Hirata2007,Bridle2007}. Following EC-B2020,  
we consider the following fiducial values for the intrinsic alignment nuisance parameters: $\{\mathcal{A}_{\rm IA},\eta_{\rm IA},\beta_{\rm IA}\}=\{1.72, -0.41, 2.17\}$.

With respect to GCph, one of the primary sources of uncertainty is the relation between the galaxy distribution and the underlying matter distribution, that is the galaxy bias. We consider a linear galaxy bias where the galaxy distribution $\delta_g$ is proportional to the matter distribution $\delta_m$,
\begin{equation}
    \delta_g(z)=b(z)\delta_m(z)\,,
\end{equation}
and the galaxy bias $b$ only depends on redshift. We note that a linear galaxy bias is sufficiently accurate to analyse large scales \citep{2021arXiv210513549D}, while non-linear galaxy bias models are needed for the very small scales \citep[see e.g.,][]{Sanchez2017,Desjacques2018}. For simplicity and in order not to mix the impact of the SSC with a non-linear galaxy bias modelling, we use the linear galaxy bias approximation in the following.

In more detail, and according to the approach used in EC-B2020, 
we consider a linear galaxy bias with a constant amplitude in each true redshift bin, that is
\begin{equation}
    b(z_i\leq z < z_{i+1})=b_i\,,
\end{equation}
where $z_i$ and $z_{i+1}$ stand for the boundaries of the $i$th redshift bin in true redshift. We choose a fiducial for the 10 galaxy bias nuisance parameters given by $b_i=\sqrt{1+\bar{z}_i}$, where $\bar{z}_i$ is the mean redshift value of each redshift bin in true redshift.

For the full analysis, taking into account the correlations between GCph and WL, we consider both a Gaussian covariance alone and its combination with the SSC. The Gaussian covariance, accounting for all correlations between angular scales, redshift combinations, and different observables, can be expressed as:
\begin{align}\label{eq:cov_cl}
    {\rm Cov}_\mr{G}&\left[C_{ij}^{AB}(\ell),C_{kl}^{CD}(\ell')\right]=\nonumber\\
    =&\frac{\delta_{\ell\ell'}^{\rm K}}{(2\ell+1)f_{\rm SKY}\Delta \ell}\left\{\left[C_{ik}^{AC}(\ell)+N_{ik}^{AC}(\ell)\right]\left[C_{jl}^{BD}(\ell')+N_{jl}^{BD}(\ell')\right]\right.\nonumber\\
    &+\left.\left[C_{il}^{AD}(\ell)+N_{il}^{AD}(\ell)\right]\left[C_{jk}^{BC}(\ell')+N_{jk}^{BC}(\ell')\right]\right\}\,,
\end{align}
where $A,B,C,D$ stand for WL and GCph, $i,j,k,l$ run over all tomographic bins, $\delta_{\ell\ell'}^{\rm K}$ represents the Kronecker delta of $\ell$ and $\ell'$, and $\Delta \ell$ stands for the width of the multipole bins. We assume that $\Delta\ell$ is large enough so that the $f_\mr{SKY}$ approximation is valid, as shown by \cite{Hivon2002}. The noise terms $N_{ij}^{AB}(\ell)$ are given by $\sigma_{\epsilon}^2\delta_{ij}^{\rm K}/\bar{n}_i$ for WL, where the variance of observed ellipticities is $\sigma_{\epsilon}^2$, and $\delta_{ij}^{\rm K}/\bar{n}_i$ for GCph. We assume that the Poisson errors on WL and GCph are uncorrelated, yielding a null noise for XC.

We consider in the following the optimistic scenario presented in EC-B2020 
concerning the multipole cuts used in the analysis. That is, we include all multipoles ranging from $\ell=10$ to $\ell=5000$ for WL and all multipoles ranging from $\ell=10$ to $\ell=3000$ for GCph and the XC terms. We note that we consider this optimistic case entering deeply into the nonlinear regime to study the impact of the SSC, where it is more relevant.

\subsection{Cosmological models}

When studying the impact of a partial-sky approach in the SSC, we consider a spatially flat Universe with cold dark matter and dark energy. We use the standard CPL parameterisation for the dark energy equation of state \citep{Chevallier2001,Linder2005}:
\begin{equation}\label{eq:cpl}
w(z)=w_0+w_a\frac{z}{1+z}\,.
\end{equation}

In addition to the $w_0$ and $w_a$ parameters describing dark energy, the cosmological model is described by the total matter density today, $\Omega_{m}$, the dimensionless Hubble constant, $h$, the baryon density today, $\Omega_{b}$, the slope of the primordial power spectrum, $n_{s}$, and the root-mean-square (RMS) of matter fluctuations on spheres of $8\,h^{-1}\,\mathrm{Mpc}$ radius, $\sigma_8$. We further assume dark energy as a minimally-coupled scalar field with sound speed equal to the speed of light and no anisotropic stress. We, therefore, neglect dark energy fluctuations and use the Parametrised Post-Friedmann (PPF) framework \citep{PPFref}, which allows the dark energy equation of state to cross $w(z)=-1$ without developing instabilities in the perturbation sector.

We consider the following set as fiducial values for our cosmological parameters:
\begin{align}\label{eq.paramsfid}
\vec{p}&=\{\Omega_{m},\,\Omega_{b},\,w_0,\,w_a,\,h,\,n_{s},\,\sigma_8,\}\nonumber\\
&=\{0.32,\,0.05,\,-1.0,\,0.0,\,0.67,\,0.96,\,0.816\}\,,
\end{align}
and we fix the sum of neutrino masses to $\sum m_{\nu}=0.06\,$eV, with one massive neutrino and two massless neutrinos.

\subsection{Derived forecast quantities}

To gauge the impact of the SSC computation on the survey's forecast statistical power, we will use two metrics. First, the signal-to-noise ratio (SNR) of the angular power spectrum of a given probe, which quantifies the strength of detection of the angular power spectrum in a model-independent way
\be
\left(S/N\right)^2 = \sum_{i,j,k,l}\sum_{\ell,\ell'} C^{AB}_{ij}(\ell)\ {\rm Cov}\left[C_{ij}^{AB}(\ell),C_{kl}^{CD}(\ell')\right]^{-1}\ C^{CD}_{kl}(\ell'),
\ee
where ${\rm Cov}$ is the (total) covariance matrix of the power spectrum, consisting of the sum of the Gaussian and SSC contributions : ${\rm Cov}={\rm Cov}_\mr{G}+{\rm Cov}_\mr{SSC}$. 

To quantify the impact on cosmological constraints, we use a second metric, the Fisher matrix
\be
\label{eq:fisher_matrix}
F_{\alpha,\beta} = \sum_{i,j,k,l}\sum_{\ell,\ell'} \frac{\partial C^{AB}_{ij}(\ell)}{\partial \theta_\alpha} \ {\rm Cov}\left[C_{ij}^{AB}(\ell),C_{kl}^{CD}(\ell')\right]^{-1} \ \frac{C^{CD}_{kl}(\ell')}{\partial \theta_\beta},
\ee
where $\theta_\alpha$ and $\theta_\beta$ are two model parameters, such as the two parameters of the dark energy equation of state. From the Fisher matrix we can derive several quantities to quantify the constraints on cosmological parameters :
\begin{itemize}
    \item The marginalised error on a given parameter $\theta_\alpha$ is given by
    \be
    \sigma_\alpha = \sqrt{\left(F^{-1}\right)_{\alpha,\alpha}}\,.
    \ee
    This expression implies that all the other parameters are marginalised over, meaning that their variation is taken into account when estimating the error.
    \item Instead, one could consider the unmarginalised error
    \be
    \sigma_\alpha^U = \sqrt{1/F_{\alpha,\alpha}}\,,
    \ee
    which is like effectively fixing the other parameters to their fiducial values. We will see in the following that this distinction is important to understand how the difference in $S/N$ between full-sky and partial-sky translates to the Fisher forecast.
    \item By considering the marginalised Fisher submatrix, $F_{w_0, w_a}$, of the Dark Energy parameter plane $(w_0, w_a)$, we can define the Dark Energy Figure of Merit as
    \be
        \mathrm{FoM}_{w_0, w_a} = \sqrt{\mathrm{det}(F_{w_0, w_a})}. 
    \ee
    This quantity is proportional to the inverse of the area delimited by the $2\sigma$ contour in the marginalised 2 parameters plane.
\end{itemize}

Finally, it is interesting to note that the SNR is a particular case of the Fisher metric for a scaling parameter $A$ defined as $C(\ell; A) = A \times C(\ell)^\mr{fid}$, where $C(\ell)^\mr{fid}$ is the angular power spectrum computed in the fiducial cosmology. 



\section{Results: impact on the statistical power}\label{Sect:results}

In this section, we compute the SNR and Fisher matrix, for the Gaussian and total (Gaussian + SSC) covariance. We focus on varying the survey masks (and $f_\mr{SKY}$) for the partial-sky (and full-sky) computation of the SSC, in order to assess the impact of survey geometry on parameter inference. First, we vary the size of the survey, then we consider different mask geometries for a fixed survey area and finally we study the dependence of the SSC computations with the survey $n(z)$. In order to simplify the notation, we denote the full-sky and partial-sky computation as fsky and psky, respectively.

\subsection{Survey Area}\label{Sect:surveysize}

In this section, we study the impact of the SSC and its implementation through the fsky and psky recipes, with respect to the suvey area. For the psky computation we consider a circular mask with a $f_\mr{SKY}$ corresponding to an area ranging from 5 to 15~000 deg$^2$. We also gauge the importance of the SSC by comparing it to the Gaussian only covariance.

\begin{figure}[!b]
\centering
\includegraphics[width=.9\linewidth]{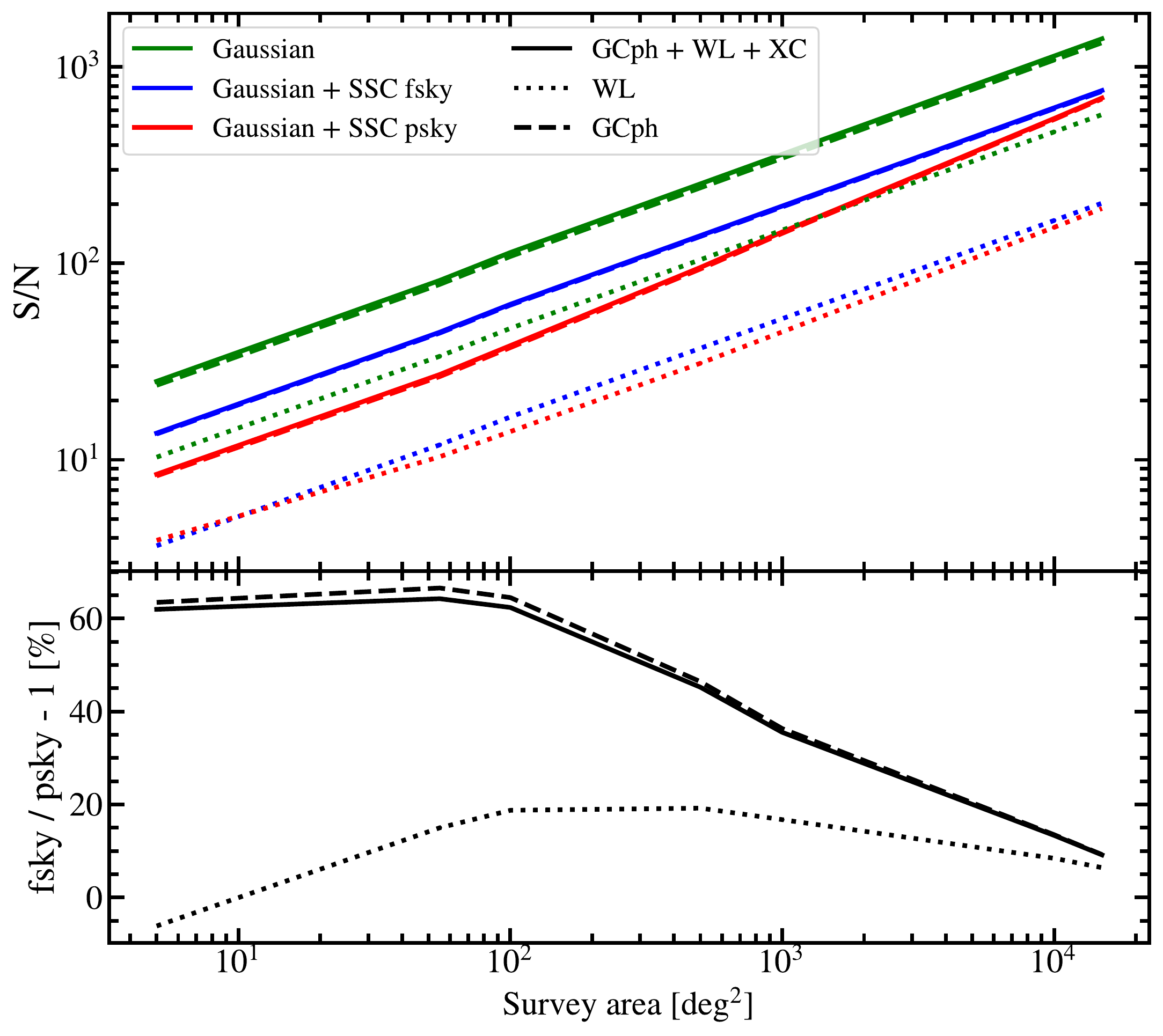}
\caption{\textit{Top:} SNR of the angular power spectrum as a function of the survey area for WL (short dashes), GCph (long dashes) and GCph+WL+XC (plain line). The $f_\mr{SKY}$ approximation is shown in blue while the partial sky computation which accounts for the mask is shown in red, the Gaussian case is also shown in green. \textit{Bottom:} relative difference in \% between the blue and the red lines, that the errors resulting from the use of the $f_\mr{SKY}$ approximation.}
\label{Fig:SNR}
\end{figure}

\begin{table*}[!htb]
    \centering
    \caption{Dark Energy Figure of Merit for all three probes when considering the Gaussian covariance or the full Gaussian+SSC covariance in fsky and psky. We show the results for the lowest and largest survey area.}
		\begin{tabular}{l | c | c c c | c}
			\hline
            \multicolumn{1}{l |}{Probe} & {Survey area [$\mr{deg}^{2}$]} & Gaussian & Gaussian + fsky SSC & Gaussian + psky SSC & fsky/psky - 1 [\%]\\
			\hline
			\hline
			\multicolumn{1}{l |}{WL}&{$5$} & $ 0.014 $ & $0.009$ & $0.012$ & $-30.87$\\
			\multicolumn{1}{l |}{}&{$15\,000$} & $43.120$ & $26.329$ & $26.335$ & $-0.02$\\
			\hline
			\hline
			\multicolumn{1}{l |}{GCph}&{$5$} & $ 0.035 $ & $0.029$ & $0.029$ & $-1.56$\\
			\multicolumn{1}{l |}{}&{$15\,000$} & $103.714$ & $88.636$ & $88.455$ & $0.20$\\
			\hline
			\hline
			\multicolumn{1}{l |}{GCph+WL+XC}&{$5$} & $0.346$ & $0.150$ & $0.166$ & $-9.58$\\
			\multicolumn{1}{l |}{}&{$15\,000$} & $1038.132$ & $454.590$ & $460.510$ & $1.30$\\
			\hline
		\end{tabular}
		\label{Tab:fom}
    \end{table*}

First, we look at how the SNR of the angular power spectrum of the probes considered in this article, WL, GCph, and GCph+WL+XC, evolves with the size of the survey. As stated in section 2, for overlapping redshift bins, the structure of the $S_{ijkl}$ matrix is more complex, so to interpret the impact of fsky and psky on our observables, it is better to compare their SNR. We show the results on Fig.~\ref{Fig:SNR}, for the psky derivation, the fsky approximation, and the Gaussian covariance case. We see that all probes are significantly affected by the SSC regardless of the size of the survey. However, the relative difference between fsky and psky depends on the probe considered and the survey area. For WL, and for a small survey of 5 deg$^2$, the fsky approximation underestimates by less than 10\% the SNR with respect to the more accurate psky computation. For larger surveys, fsky systematically overestimates the SNR, with a maximum of 20\% for a 100 deg$^2$ survey and a minimum of 5\% for a 15\,000 deg$^2$ survey. For GCph+WL+XC the behaviour of the SNR is similar to the GCph only case, since the GCph dominates the signal with respect to WL. For these two probes fsky leads to an overestimate of the SNR regardless of the survey area. The relative difference is maximum, with a 60\% overestimate, for small surveys between 5 and 100 deg$^2$ and reaches a minimum of $\sim$ 7\% for 15\,000 deg$^2$, close to the WL case.

To understand how these results translate in terms of parameter constraints, we perform a Fisher forecast analysis following Sect.~\ref{Sect:forecasts}.
Table \ref{Tab:fom} gathers the values of the Dark Energy FoM for all probes, considering the Gaussian covariance and the total Gaussian+SSC covariance in fsky and psky, for survey areas of 5 and 15~000 deg$^{2}$. As expected from the SNR, the SSC has a large impact on the Dark Energy FoM, especially for WL and GCph+WL+XC, for which it is reduced by half for both low and large survey area. For low survey areas, the fsky approximation underestimates the FoM, especially for auto-correlations of probes, the most impacted by the SSC. However, for the largest survey areas, the difference between fsky and psky is almost negligible.

\begin{figure}[!h]
\centering
\begin{subfigure}{0.46\textwidth}
\includegraphics[width=.9\linewidth]{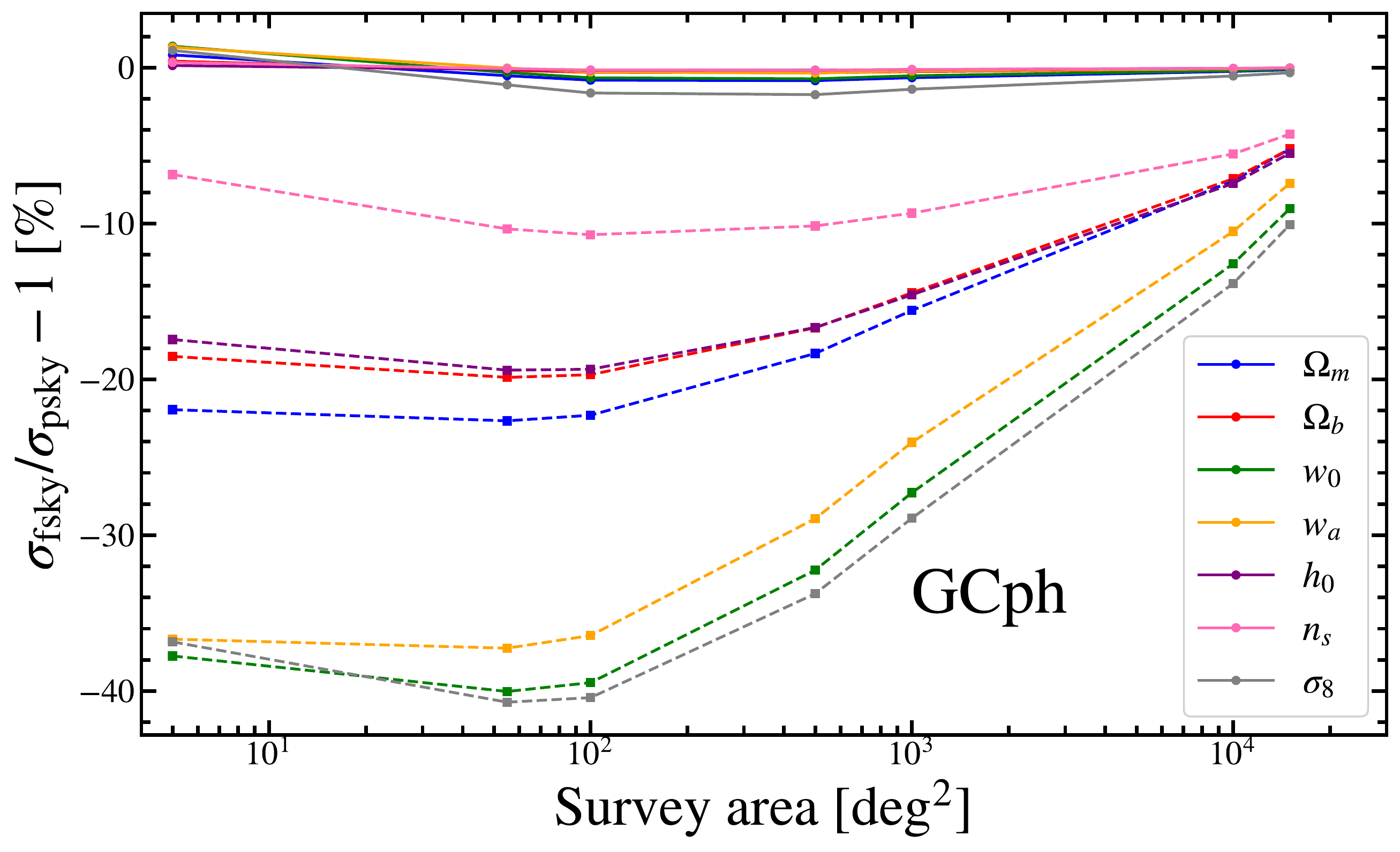}
\caption{} 
\label{fig:GC_area}
\end{subfigure}
\begin{subfigure}{0.46\textwidth}
\includegraphics[width=.9\linewidth]{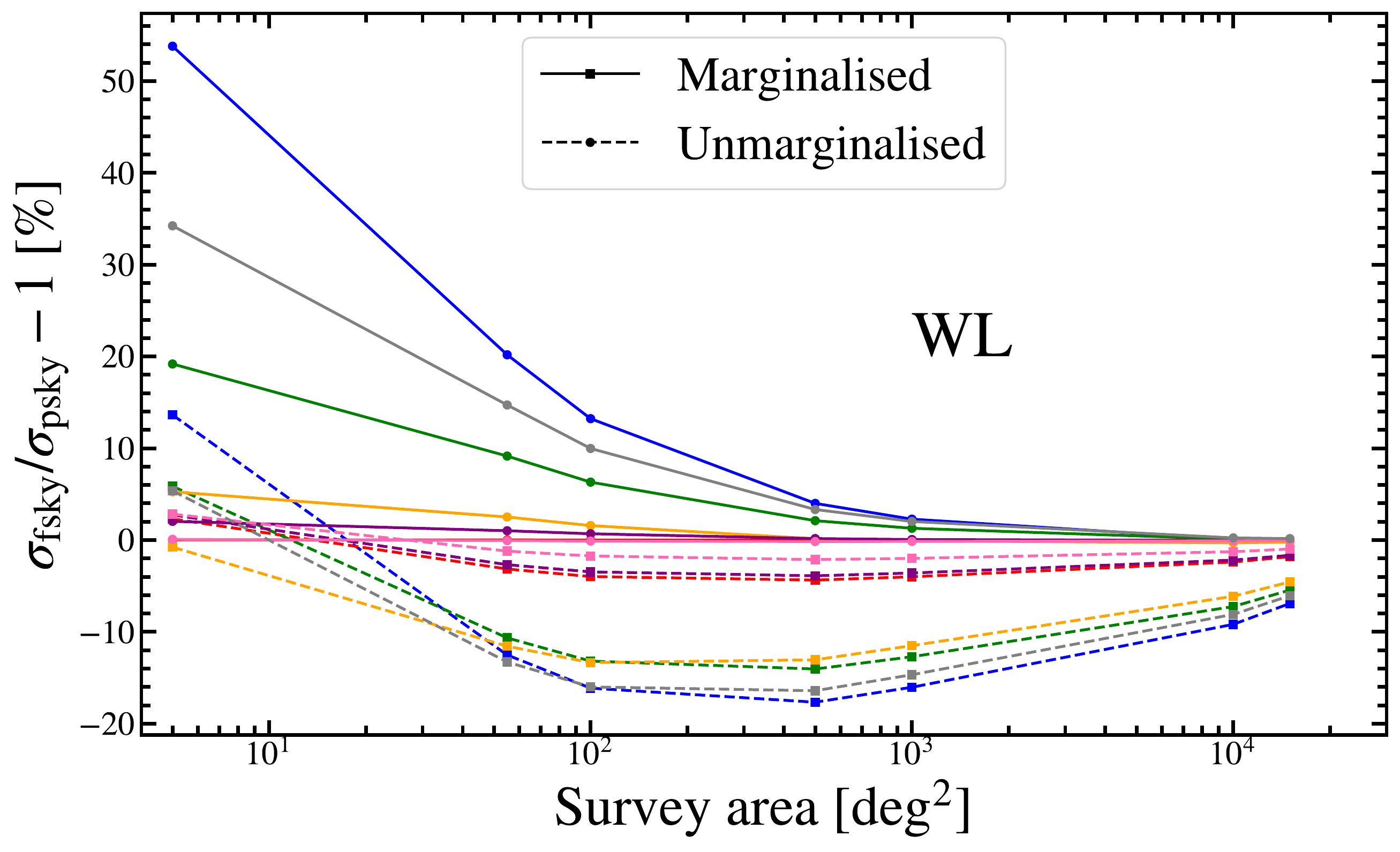}
\caption{} 
\label{fig:WL_area}
\end{subfigure}
\begin{subfigure}{0.46\textwidth}
\includegraphics[width=.9\linewidth]{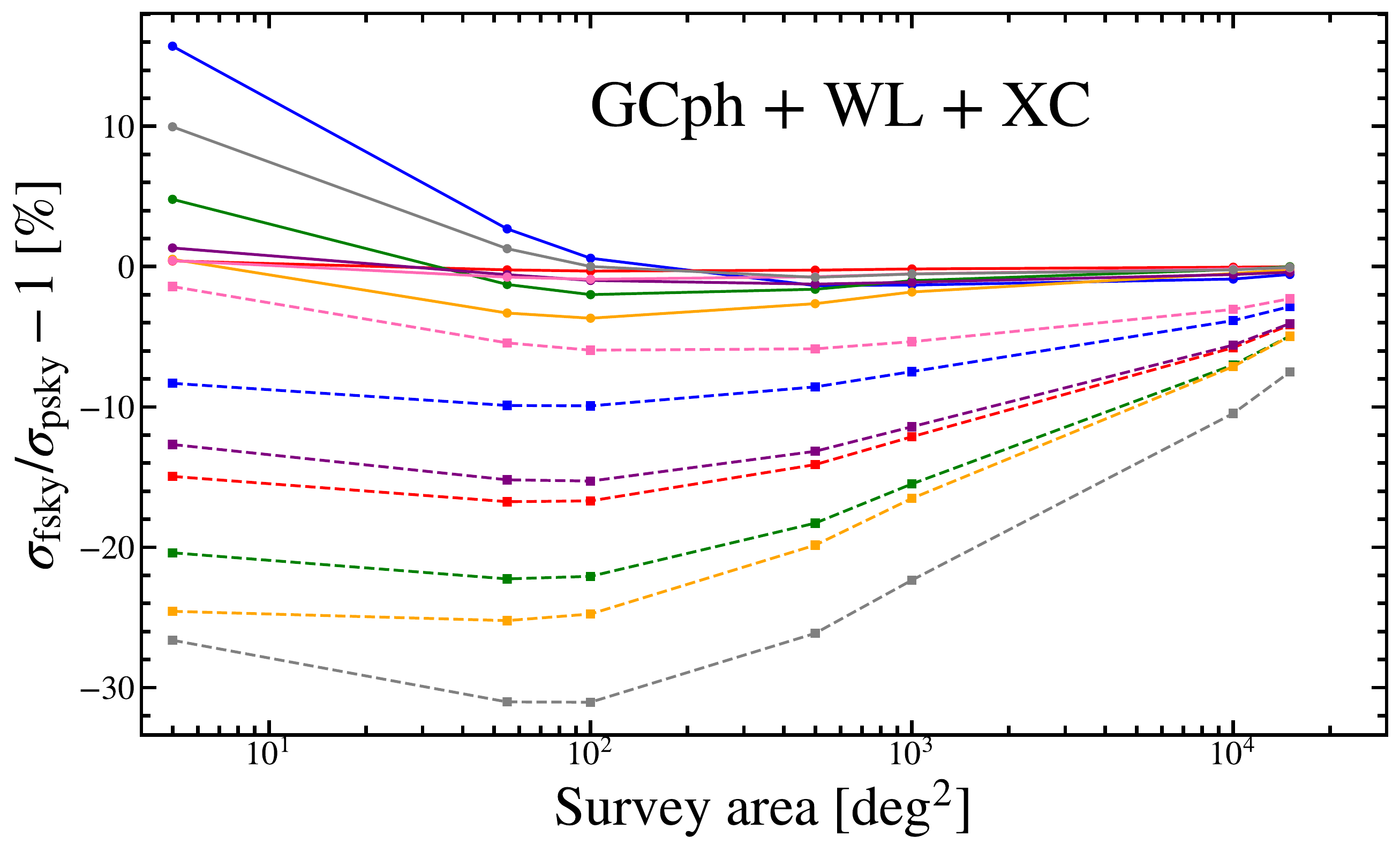}
\caption{} 
\label{fig:XC_area}
\end{subfigure}
\caption{Forecast errors on cosmological parameters, for the Gaussian+SSC covariance, using different area of  circular masks for GCph (a), WL (b), and GCph+WL+XC (c). For the three panels, we show the relative difference in \% between fsky and psky for each cosmological parameter. The plain lines corresponds to the constraints when marginalising on all parameters (cosmological and nuisance) and the dashed lines when there is no marginalisation.}
\label{Fig:area}
\end{figure}

To further understand these results, we present in Fig.~\ref{Fig:area} the marginalised and unmarginalised constraints on all cosmological parameters and survey area between 5 deg$^2$ and 15\,000 deg$^2$. In this figure, we consider the Gaussian+SSC covariance using the psky derivation or the fsky approximation, for different probes. In the unmarginalised case, the error bars resulting from using the fsky approximation or the psky computation follow the same evolution with the survey area as did the SNR. That is, for all probes, when the fsky approximation leads to an overestimated SNR, it also results in underestimated error bars on cosmological parameters, and conversely. Interestingly, when marginalising on all the varied parameters, that is cosmological and nuisance parameters, these results change. For GCph (see Fig~\ref{fig:GC_area}), the difference between fsky and psky is largely reduced, giving a relative difference between 1 and $-2$\% for all cosmological parameters and survey area. For WL (Fig~\ref{fig:WL_area}), the marginalisation has an opposite effect and the error is overestimated with fsky, especially for the smallest surveys, going up to 50\% increase for $\Omegam$, 35\% for $\sigma_{8}$ and 20\% for $w_0$, in the case of a 5 deg$^2$ survey. The relative difference is below 10\% for surveys larger than 100 deg$^2$, and goes close to zero for the largest areas. For GCph+WL+XC, Fig~\ref{fig:XC_area}, the situation is the same as for WL, with a smaller amplitude of the relative difference for $\Omegam$, $\sigma_{8}$ and $w_0$, except for $w_a$ which shows a small underestimation of its error, $\sim-2\%$, on intermediate survey areas.
An interesting point to note is that the most impacted cosmological parameters are the ones related to the amplitude of the power spectrum, namely $\Omegam$, $\sigma_{8}$, $w_0$, and $w_a$. This result is expected because the effect of the SSC appears on the amplitude of the power spectrum through the change in the background density $\delta_b$.

\begin{figure}
\centering
\includegraphics[width=.95\linewidth]{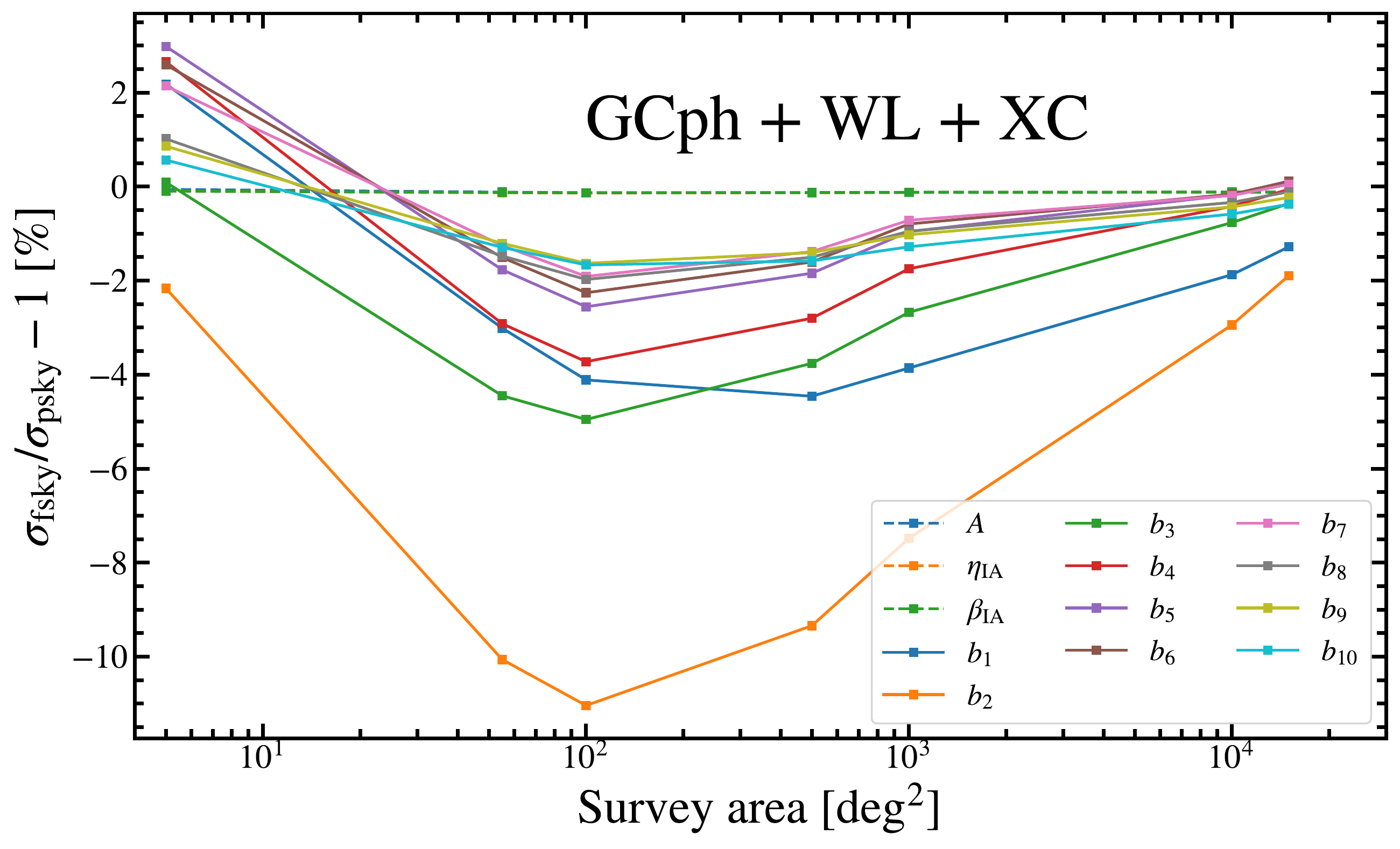}
\caption{Marginalised errors on nuisance parameters using circular masks of different areas for GCph+WL+XC. We show the relative difference in \% between fsky and psky for each nuisance parameter. The nuisance parameters associated to WL are shown in dashed lines and the ones associated with GCph in plain lines.}
\label{Fig:XC_nuisance}
\end{figure}

Overall, we see that, for the marginalised constraints, the complete treatment of the mask in the derivation of the SSC is not necessary for large areas, representative of upcoming stage-IV cosmological surveys. However, marginalising has an important effect on the impact of the SSC. Despite the fact that the SNR resulting from the fsky or psky computations largely differ, for GCph, the difference is seemingly absorbed in the nuisance parameters through marginalisation. In contrast, with WL, when marginalising, the difference is transferred to the cosmological parameters.

\begin{figure*}
\centering
\includegraphics[width=1.\linewidth]{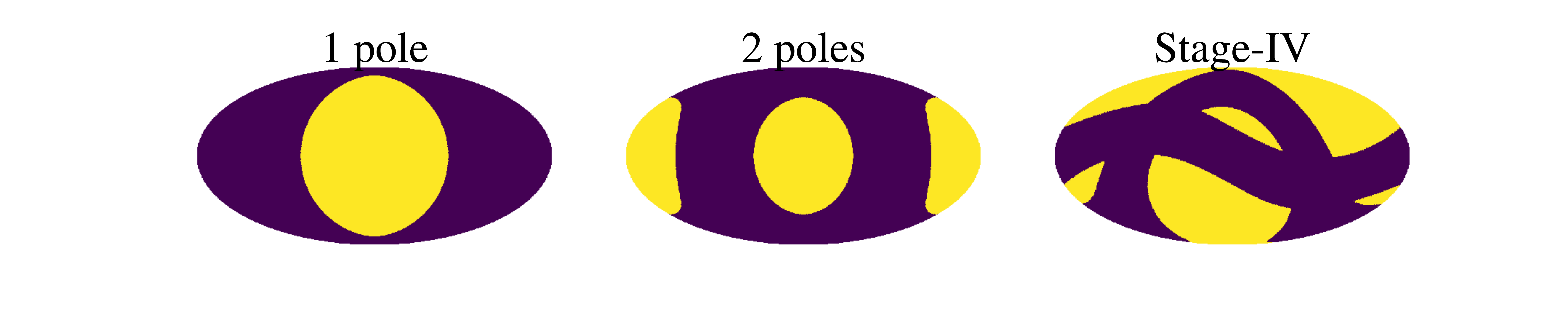}
\caption{Mollweide view of the three masks considered in Sect.~\ref{Sect:surveygeometry}. They all have an area of 15~000 deg$^2$, corresponding to $f_\mr{SKY} = 0.364$. The yellow area is the observable region.}
\label{Fig:masks}
\end{figure*}

To confirm this interpretation, we look at the forecasted errors on all nuisance parameters, that are the ten constant galaxy biases $b_i$ and the three intrinsic alignment parameters $\mathcal{A}_{\rm IA}$, $\eta_\mr{IA}$ and $\beta_\mr{IA}$, when accounting for the SSC with fsky or psky. The results are shown in Fig~\ref{Fig:XC_nuisance} for GCph+WL+XC. The galaxy biases are showing different constraints depending on the SSC recipe, with a maximum negative relative difference between -2 and -5\% for all $b_i$, except for $b_2$ which goes down to -10\%. On the other hand, the errors on the intrinsic alignment parameters do not change whichever recipe is used.

This is due to the fact that the simple model we use for the galaxy bias is just an amplitude factor on the power spectrum. The $b_i$ nuisance parameters thus absorb, through the marginalisation, the effect of SSC, which also modulates the amplitude of the power spectrum through the variation of the background density $\delta_b$. The difference of SSC between fsky and psky is then mostly transferred to the linear galaxy bias, while the IA parameters are insensitive to SSC. We observe the same behaviour for GCph and WL alone. Interestingly, \cite{Wadekar2020} found a similar effect of the marginalisation when accounting for the full non-Gaussian covariance in a spectroscopic GC analysis.    


The above results can be summarised in two important points. First, the SNR is a misleading metric to quantify the impact of a correct psky treatment of the SSC. Second, even if the difference on marginalised cosmological constraints is close to zero between the fsky and psky approaches for large survey areas, for unmarginalised constraints the difference can be of the order of 10\%. 
Since the difference can be absorbed by the nuisance parameters, accounting for the full geometry of the mask when computing the SSC will be essential when tight priors on nuisance parameters are included in the analysis.

\subsection{Survey geometry}\label{Sect:surveygeometry}

\begin{figure}[t!]
\centering
\begin{subfigure}{0.5\textwidth}
\includegraphics[width=.9\linewidth]{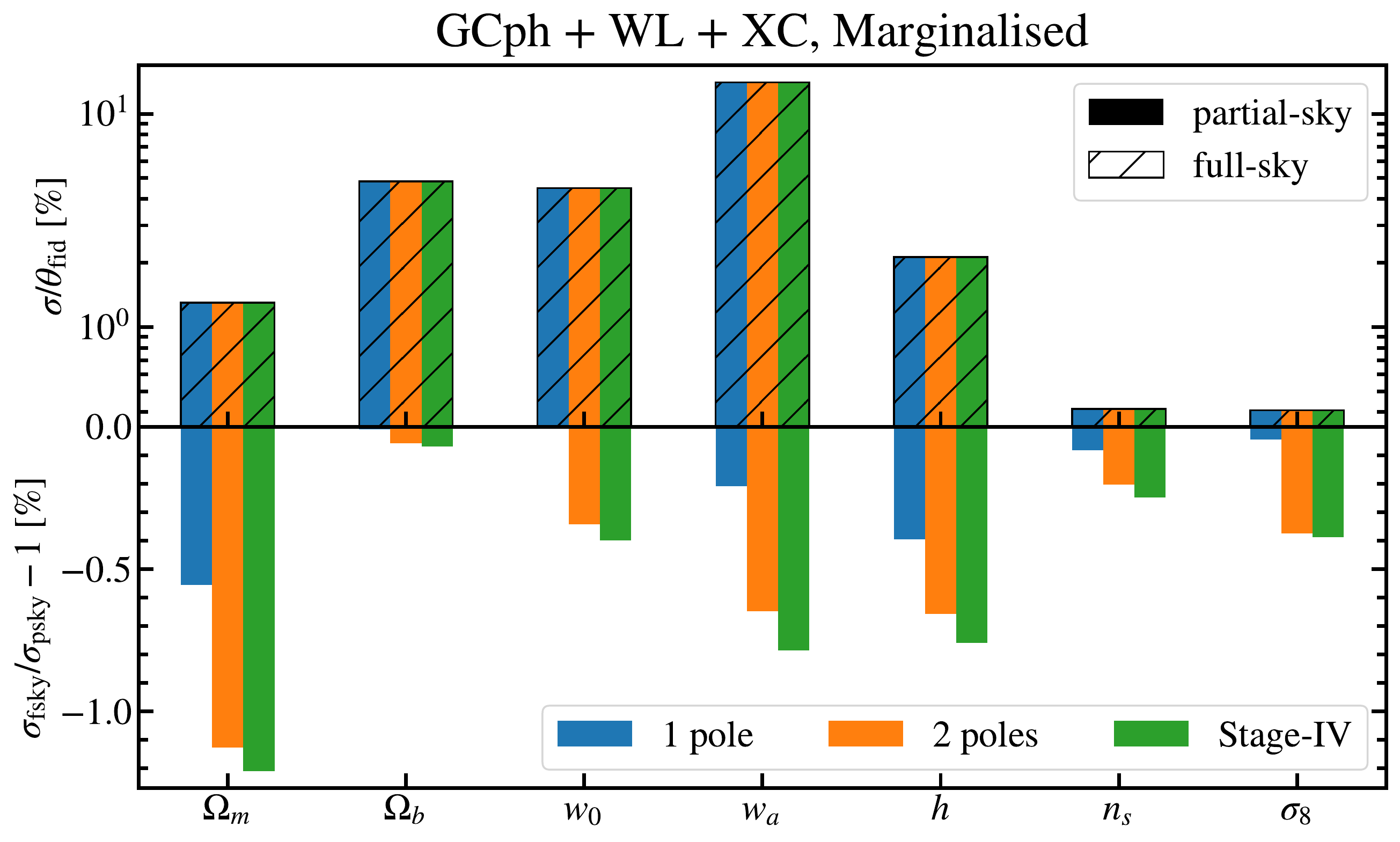}
\caption{} 
\label{fig:XC_geometry_marg}
\end{subfigure}
\begin{subfigure}{0.5\textwidth}
\includegraphics[width=.9\linewidth]{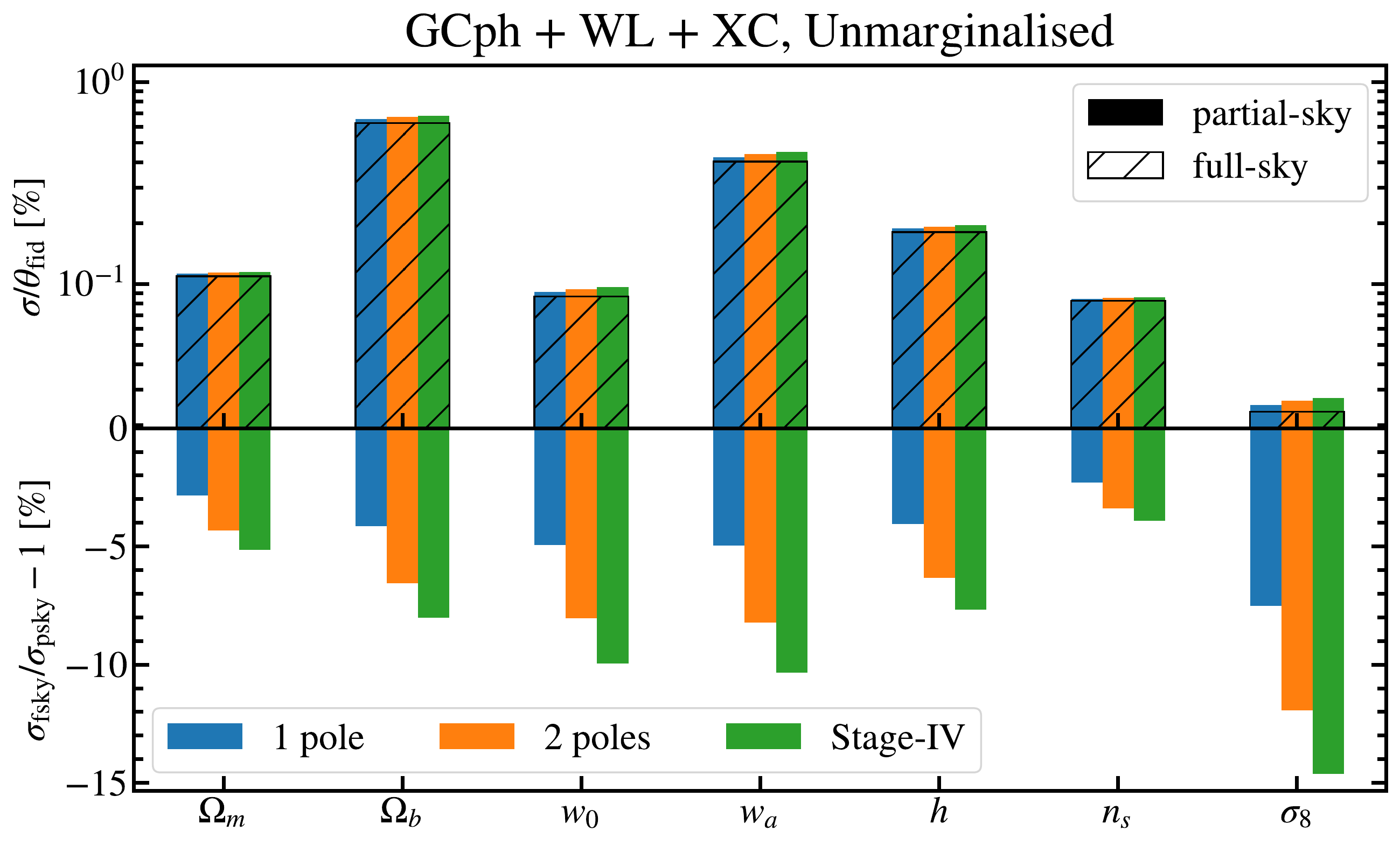}
\caption{} 
\label{fig:XC_geometry_unmarg}
\end{subfigure}
\caption{Forecast errors on cosmological parameters for different mask geometry and a 15~000~deg$^2$ survey, for GCph+WL+XC, when marginalising (a) or not (b). \textit{Top}: Relative error in \% for each cosmological parameter. The filled coloured bars represent the constraints obtained with the psky derivation, using a circular mask (blue), a mask divided in 2 circles (orange) and a stage-IV-like mask (green). The hatched empty bars are the ones obtained in the fsky approximation. \textit{Bottom}: Relative difference between the fsky and the psky standard deviation for each mask geometry.}
\label{Fig:geometry_histo}
\end{figure}

In this section, we study the impact of the two different SSC recipes with respect to the geometry of the survey. We consider a survey with an area of 15~000~deg$^2$ and three different geometries: A single circular patch (such as the ones used in Sect.~\ref{Sect:surveysize}) -- dubbed 1 pole, two separated circular patches -- dubbed 2 poles, and a survey with a geometry close to future stage-IV surveys such as Euclid, where the galactic and zodiacal plans have been removed. These three masks are represented in a Mollweide view in Fig.~\ref{Fig:masks}. 

We perform a Fisher forecast in the same setting as described in Sect.~\ref{Sect:forecasts} for the three masks considered. In Fig~\ref{Fig:geometry_histo}, we present the resulting marginalised and unmarginalised constraints in the psky and fsky cases for GCph+WL+XC.
For the marginalised constraints (Fig~\ref{fig:XC_geometry_marg}), as we already saw in the previous section, the relative difference between fsky and psky is close to zero for the simplest mask geometry. A more complex mask geometry leads to larger discrepancies between the two approaches, which however remain smaller than or close to 1\% for all cosmological parameters. In the unmarginalised case (Fig~\ref{fig:XC_geometry_unmarg}), we can see the same effect that was discussed in the previous section: The relative difference is larger when we do not marginalise. Additionally, regarding the impact of the geometry, we observe the same trend as in the marginalised constraints: The difference increases with the complexity of the mask. For the most complex stage-IV mask, the fsky approximation underestimates the error by almost 10\% for $w_0$, $w_a$, and 15\% for $\sigma_8$, in contrast to a 5\% difference observed with the 1 pole mask. Similar results are obtained with GCph and WL alone.

Therefore, for large, stage-IV-like survey areas, the marginalised errors do not depend strongly on the mask geometry. However, as discussed in Sect.~\ref{Sect:surveysize}, this result will not hold when adding tight priors on nuisance parameters, which is equivalent to the unmarginalised errors case.

\subsection{Survey galaxy distribution}\label{Sect:surveynz}

\begin{figure}[t]
\centering
\includegraphics[width=.9\linewidth]{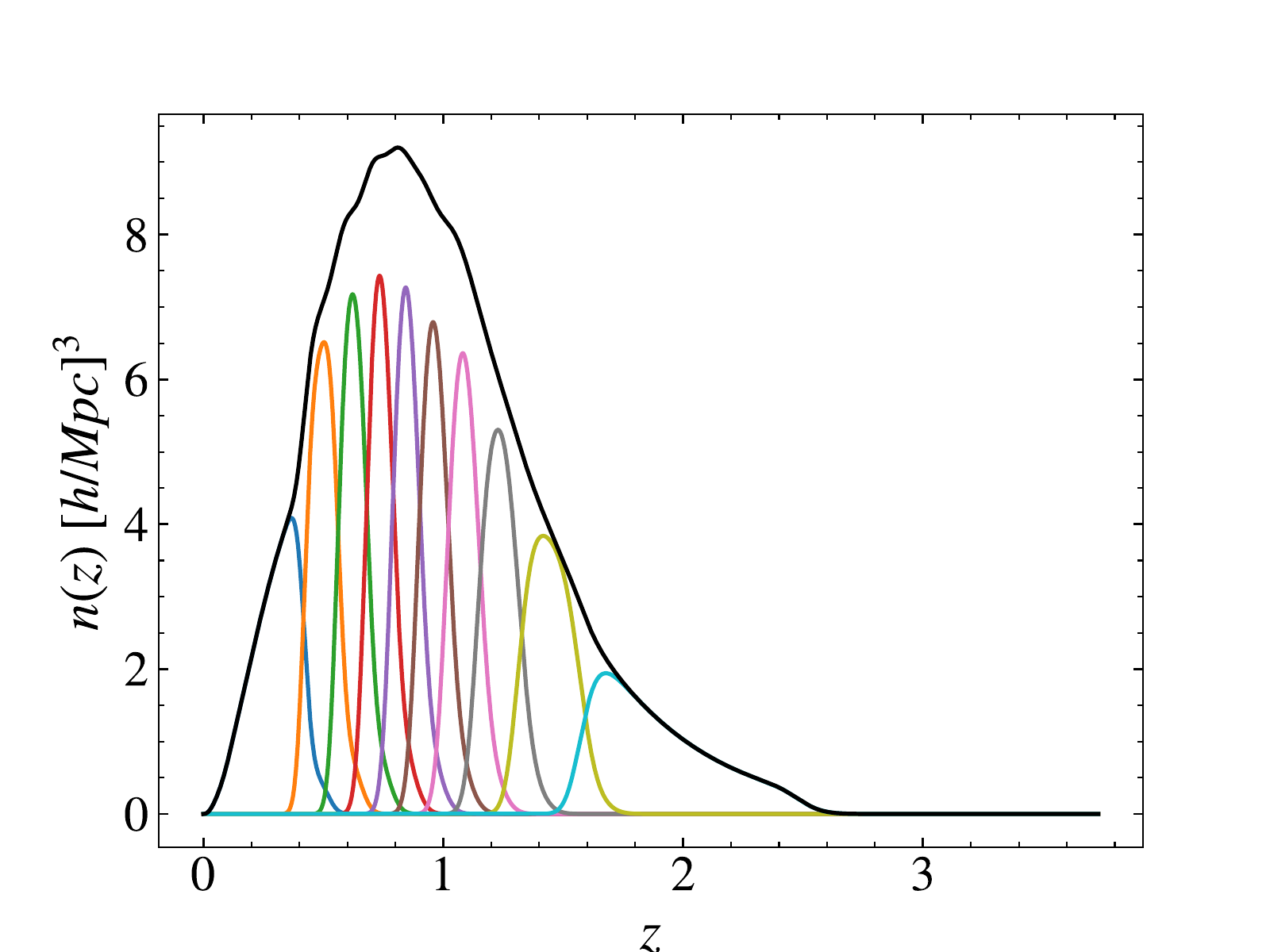}
\caption{Normalised galaxy number density distribution in the 10 photometric redshift bins, in the tight case. The black line is the sum of all the redshift bins.}
\label{Fig:lsst_nz}
\end{figure}

In this section, we study the impact of the two different SSC recipes, comparing the forecasts obtained for different galaxy distribution $n(z)$. Indeed, since the $S_{ijkl}$ matrix is computed from integrals of the kernels over $z$, it depends on the input $n(z)$. We use the forecast results from the previous sections and compare them to constraints obtained when changing some of the parameters of the assumed galaxy distribution (see Eq.~\eqref{eq:nz1}, \eqref{eq:nz2} and \eqref{eq:nz3}). We consider the same $n(z)$ as before and a new $n(z)$ with tighter redshift bins. The new $n(z)$ -- dubbed 'tight', is shown in Fig.~\ref{Fig:lsst_nz}, whereas the original $n(z)$ -- dubbed 'wide', can be seen on Fig.~\ref{Fig:euclid_nz}. In addition, we also consider a third $n(z)$, similar to the wide one, but with a different outlier fraction $f_{out} = 0.25$, while it was 0.1 in the previous section. Indeed, the fraction of outlier redshifts is one of the most important issue to handle for photometric surveys. The wide $n(z)$ with $f_{out} = 0.25$ is not shown as it is visually very similar to the case with $f_{out} = 0.1$, but the interested reader can find in Appendix \ref{Appendix_nz} a table with the values of the different parameters used to compute the three $n(z)$ considered in this section. Here, we use the 1 pole, 15\,000 deg$^2$ mask introduced in Sect.~\ref{Sect:surveygeometry} and visible on the left panel of Fig.~\ref{Fig:masks} in all three cases. 

\begin{figure}[h]
\centering
\begin{subfigure}{0.5\textwidth}
\includegraphics[width=.9\linewidth]{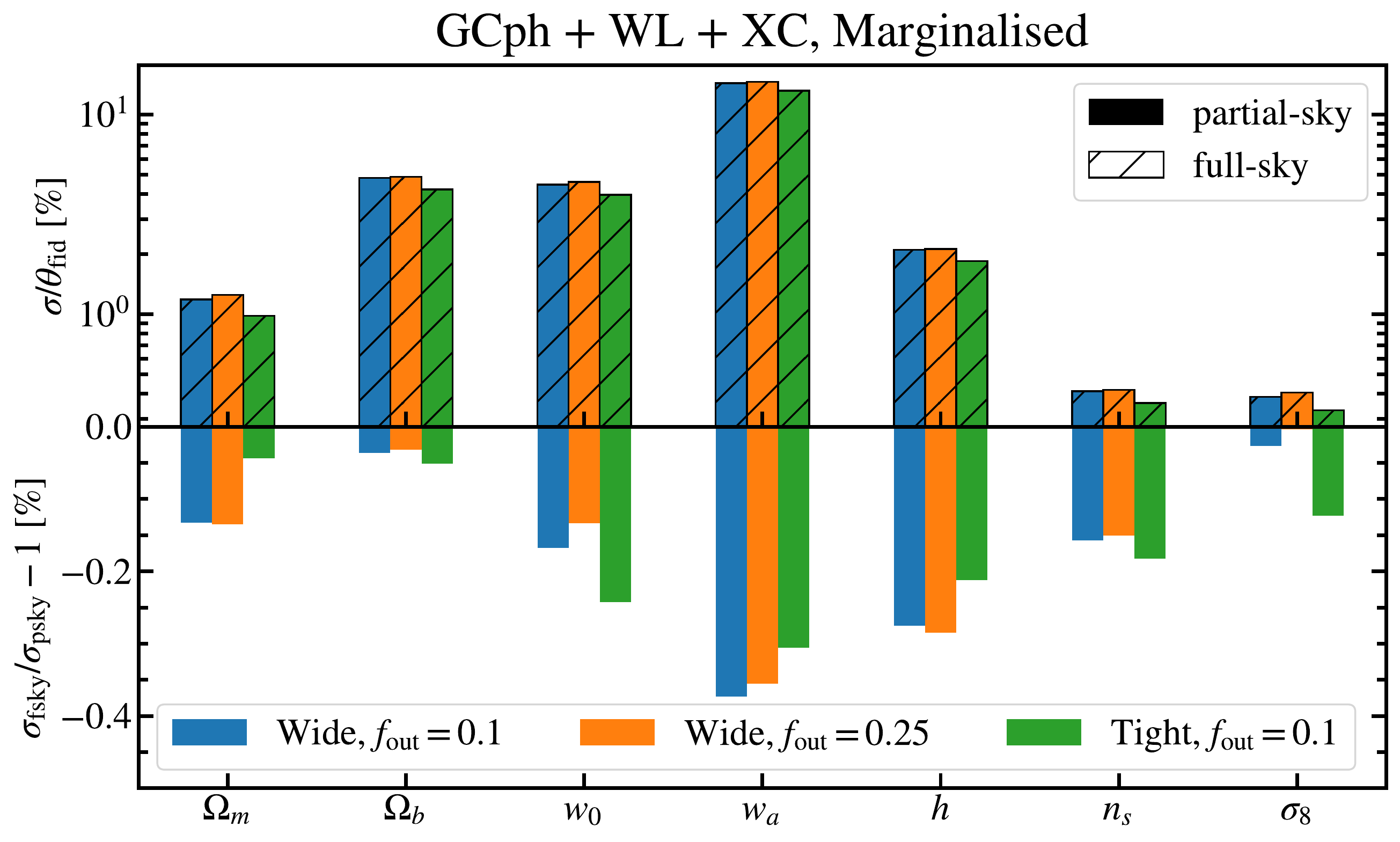}
\caption{} 
\label{fig:XC_S4_marg}
\end{subfigure}
\begin{subfigure}{0.5\textwidth}
\includegraphics[width=.9\linewidth]{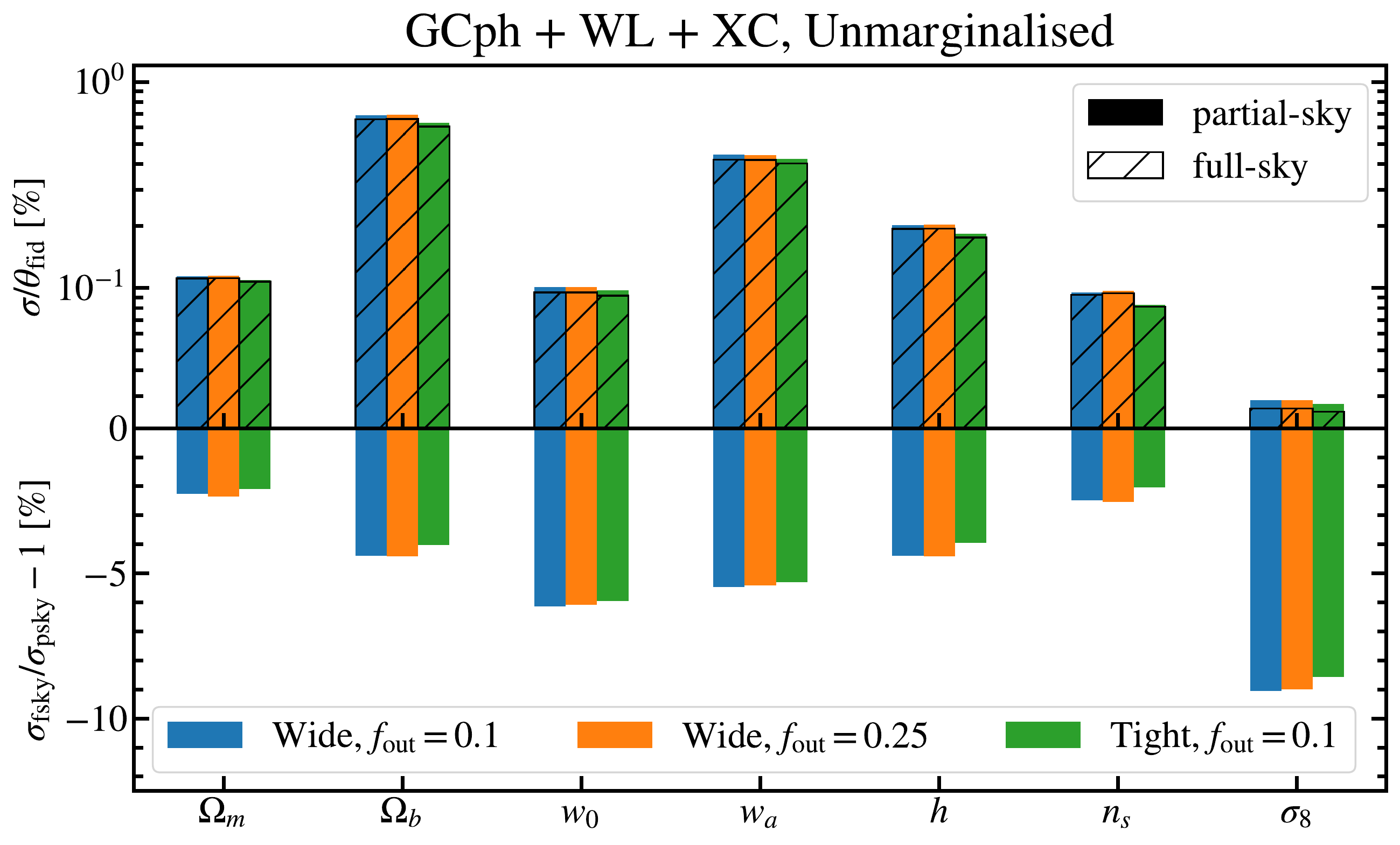}
\caption{} 
\label{fig:XC_S4_unmarg}
\end{subfigure}
\caption{Forecast errors on cosmological parameters for three $n(z)$: with wide redshift bins and an outlier fraction of $f_{out} = 0.1$ or $f_{out} = 0.25$ and with tight redshift bins. The results are shown for GCph+WL+XC, when marginalising (a) or not (b). We suppose a circular 15\,000 deg$^2$ survey in the three cases. \textit{Top}: Relative error in \% for each cosmological parameter. The filled coloured bars represent the constraints obtained with the psky derivation, using the wide $n(z)$ with $f_{out} = 0.1$ (blue) or $f_{out} = 0.25$ (orange) and the tight $n(z)$ (green). The hatched empty bars are the ones obtained in the fsky approximation. \textit{Bottom}: Relative difference between the fsky and the psky standard deviation for each galaxy distribution.}
\label{Fig:S4_histo}
\end{figure}

We perform a Fisher forecast in the same setting as described in Sect. \ref{Sect:surveysize} for the three $n(z)$ considered. In Fig~\ref{Fig:S4_histo}, we present the resulting marginalised and unmarginalised constraints in the psky and fsky cases for GCph+WL+XC. 
Looking at the top panels of Fig~\ref{fig:XC_S4_marg} and Fig~\ref{fig:XC_S4_unmarg}, we see that for all parameters, the relative error is larger when using the wide $n(z)$. This can be related to the amount of overlap in the different $n(z)$ functions:  for equal galaxy number densities, less overlap means less correlation between redshifts and, in turn, more independent information in each bin. In addition we saw in Sect.~\ref{Sect:method} that for the pairs of bins which do not overlap the SSC produces anti-correlations, which tends to increase the SNR compared to positive correlations. Increasing the outlier fraction to $f_{out} = 0.25$ doesn't seem to have a large impact on the constraints. 


For the marginalised constraints (Fig~\ref{fig:XC_S4_marg}), the difference between fsky and psky is sub-percent regardless of the $n(z)$ considered. Given the low relative difference, we don't find a specific trend for a galaxy distribution or another. It can be larger for the wide or the tight case depending on the parameter considered. We find the same result for WL and GCph individually. For the unmarginalised constraints (Fig~\ref{fig:XC_S4_unmarg}) however, the difference between the two SSC recipes is slightly larger in the wide case than in the tight case. This could be explained by the larger error bars obtained with the former.


\section{Conclusion}\label{Sect:conclusion}

Many works have shown that the SSC can have a large impact on cosmological constraints coming from a variety of standard cosmological probes \citep{Hu2003, Barreira2018b, Lacasa2020}. Most works focused on describing the SSC using the approximation of flat-sky \citep{DES_Y1,KrauseEifler_2017} or full-sky \citep{Lacasa2016,Lacasa2019}. A few more recent works \citep{Lacasa2018,Barreira2018b,DES_Y3} have studied the more realistic case of a survey mask. In this article, we have extended the fast SSC approximation introduced in \citet{Lacasa2019} to an accurate partial-sky treatment of the mask. This new SSC recipe retains the advantage of \citet{Lacasa2019} : it is fast to compute, mostly model-independent, and the code will be made available publicly when this article is published. In this article, we have used it to investigate the impact of the mask geometry on the SSC, and, in turn, on cosmological parameter inference for future surveys. 

Comparing the standard flat-sky and full-sky approximations to our new, more accurate, partial-sky computation, we have first found in Sect.~\ref{Sect:SSC-compare} that the flat-sky approximation gives a satisfying estimate of SSC-induced correlations between redshift bins only for survey areas smaller than 5 deg$^2$, and specifically for $z>1$. On the other hand, the full-sky approach can recover the partial-sky SSC for wide surveys larger than 15\,000 deg$^{2}$ with a $10\%$ precision, and, as expected, performs better at higher redshifts.
Following these first results, we have chosen to focus the analysis of Sect.~\ref{Sect:results} on the full-sky and partial-sky methods to explore the impact of survey geometry on cosmological constraints. We have considered the main photometric probes of the upcoming stage-IV surveys, the so-called 3x2pts analysis, composed of photometric galaxy clustering, weak lensing and their cross-correlation (also called galaxy-galaxy lensing). We have performed Fisher forecasts, following the methodology of EC-B2020 -- see Sect.~\ref{Sect:forecasts}.

First, in Sect.~\ref{Sect:surveysize}, we have considered a simple circular geometry for the mask and varied its area. Starting from a naive signal-to-noise ratio (SNR) comparison, we have seen that accounting for the SSC greatly reduces the SNR for all probes, compared to the Gaussian, covariance-only case. This result holds for the full-sky approximation, despite it overestimating the SNR, especially for small survey areas. Moving to parameter inference, we have confirmed that accounting for the SSC reduces by half the Dark Energy FoM for WL and GCph+WL+XC. Regarding the comparison between full-sky and partial-sky SSC, the impact on cosmological constraints is more challenging to interpret. The unmarginalised errors are in agreement with the SNR: The full-sky approximation leads to an underestimation of the error-bars (due to an overestimation of the SNR). After marginalisation, the difference in the errors induced by the two approximations is reduced, except for WL alone. That is because the large difference observed in the SNR is absorbed by the galaxy biases of GCph, while the intrinsic alignment parameters of WL are insensitive to the change of the SSC recipe. Overall, for both marginalised and unmarginalised constraints, the fsky approximation performs better as we increase the survey area, with a less than $10 \%$ relative difference on cosmological errors with psky for a 15~000 deg$^2$ survey.

Comparing surveys of area 15~000 deg$^2$ but different footprints, we have found in Sect.~\ref{Sect:surveygeometry} that the more complex the mask geometry, the larger the difference between full-sky and partial-sky. For the marginalised constraints this difference is negligible as it is always sub-percent. For unmarginalised errors, however, the difference cannot be neglected as for some parameters, including $w_a$, it ranges from 5\% for the simplest mask to 10\% for the most complex one.
Finally, in Sect.~\ref{Sect:surveynz}, we compared constraints obtained with a circular 15~000 deg$^2$ mask, with those obtained with the same mask but with a different galaxy number density distribution $n(z)$. For a $n(z)$ which has less overlap between redshift bins, the difference between full-sky and partial-sky in the unmarginalised constraints is found to decrease. That is because, for the same galaxy number density, a $n(z)$ with less overlap between redshift bins leads to smaller error bars. However, the difference is small and the changes in the $n(z)$'s we considered doesn't have a significant impact on the difference between full-sky and partial-sky for all probes.

Overall, these results show that, for wide surveys, such as the future stage-IV surveys and, in particular, \textit{Euclid}, a complete treatment of the angular mask geometry when estimating the SSC is not crucial, and the full-sky approximation is sufficient. However, as also reported in \cite{Wadekar2020}, the marginalisation over the nuisance parameters can absorb the effect of non-Gaussian covariance and if tight priors are applied to these parameters, the effect of the mask could be no longer negligible. 

As the partial-sky method we presented in this article is made public\footnote{Available at \url{https://github.com/fabienlacasa/PySSC}.}, it can be used for further analysis to accurately account for the SSC in any cosmological survey.


\section*{Acknowledgements}

SGB, FL, IT, MA, PB and AG thank the organisers of the Euclid France 2019 summer school\footnote{See \url{https://ecole-euclid.cnrs.fr}.}, where this project was initiated.

FL was partially supported by a postdoctoral grant from Centre National d’Études Spatiales (CNES).

IT acknowledges support from the Spanish Ministry of Science, Innovation and Universities through grant ESP2017-89838, and the H2020 programme of the European Commission through grant 776247.  

AG acknowledges financial support from the European Research Council under ERC grant number StG-638743 ("FIRSTDAWN") and from the Trottier Chair in Astrophysics, through the McGill Astrophysics Fellowship, as well as from the Canadian Institute for Advanced Research (CIFAR) Azrieli Global Scholars program  and the Canada 150 Programme. 

This work also acknowledges support from the OCEVU LABEX (Grant No. ANR-11-LABX-0060) and the A*MIDEX project (Grant No. ANR-11-IDEX-0001-02) funded by the Investissements d'Avenir french government program managed by the ANR. 

This research made use of \texttt{astropy}, a community-developed core Python package for astronomy \citep{astropy,astropy2}; of \texttt{matplotlib}, a Python library for publication quality graphics \citep{hunter_2007}; of \texttt{scipy}, a Python-based ecosystem of open-source software for mathematics, science, and engineering \citep{scipy} -- including \texttt{numpy} \citep{numpy}; of \texttt{class}, a Boltzmann solver code for cosmology \citep{Blas:2011rf} and of the {\tt Healpix} package \citep{Gorski2005}.

\bibliographystyle{aa}
\bibliography{bibliography}

\appendix

\section{Partial-sky code validation}
\label{Appendix_validation}
In this appendix we validate the outputs of the partial-sky code \texttt{PySSC.Sij\_partsky}\footnoteref{note1} allowing the computation of the $S_{ij}$ matrix (in the case of a single observable, see Sect. \ref{Sect:SSC-partsky}) for a sample of different masks and radial window function configurations. We compare these results with the corresponding numerical predictions of the \texttt{AngPow}\footnote{\label{note2}\href{https://gitlab.in2p3.fr/campagne/AngPow}{gitlab.in2p3.fr/campagne/AngPow}} \citep{Campagne:2017xps,Angpow:2018} public software.

As pointed out in Eq. (\ref{eq:4_smat_partsky}), the $S_{ijkl}$ matrix can be seen as the evaluation of the auto- and cross-angular power spectrum between observables and redshift bins (only redshift bins for the $S_{ij}$ matrix).
\texttt{AngPow} is specifically designed to compute the integral Eq. (\ref{eq:angularmonopoledefinition}) for arbitrary multipoles $\ell$, as long as the input radial window function is transformed such that $W(z) \rightarrow W^2(z)$ for a similar result between the two methods  \citep[see the \texttt{AngPow} integral in][]{Campagne:2017xps}.

For this exercise, we choose several mask settings accounting for different SSC contribution levels, including circular angular patches of respectively [$1$, $5$, $55$, $100$, $500$, $1000$, $10\,000$] deg$^2$ (corresponding to a fraction of the sky $f_{\rm SKY} = [0,002, 0.01, 0.13, 0.24, 1.21, 2.49, 24.35]\times 10^{-2}$) and a stage-IV mask (as featured in Fig. \ref{Fig:masks}) equivalent to a full sky survey without the galactic plane and the zodiacal ($15\,000$ deg$^2$ or equivalently $f_{\rm SKY} = 0.364$). For each of them we use several arbitrary Gaussian radial window functions centered at redshifts $z = [0.5,0.75,1,1.25,1.5]$ with variance $0.01$, allowing them to overlap. 

Both codes are run in the context of a linear $\Lambda$CDM cosmology ($h=0.67$ , $\Omega_b = 0.05$, $\Omega_{cdm} = 0.27$, $n_s = 0.96$, $A_s = 2.1265.10^{-9}$) with an input power spectrum computed at $z=0$ using the \texttt{CLASS} code\footnote{\href{http://class-code.net/}{class-code.net}} \citep{Blas:2011rf} in the range $k\in[10^{-5},  3 \times 10^{-1}] h/$Mpc. The required maximum $\ell$ in the computation of the $C_\ell$'s for a convergence of the variance within 5\% is computed by the \texttt{PySSC.find\_lmax} function for both codes. Here $\ell_{max} = [243,218,115,90,51,36,11]$ for the respective circular patches and $\ell_{max} = 20$ for the stage-IV one. 
Moreover, the mask spectrum $C_\ell(\mathcal M)$ (see Eq. \ref{eq:4_smat_partsky}) is numerically computed using the healpix code\footnote{\href{http://healpix.sourceforge.net}{healpix.sourceforge.net} }\citep{Zonca2019,Gorski2005}.

More technically, we use as \texttt{Angpow} settings a trapezoidal radial quadrature with associated \texttt{radial\_order} parameter set at 350 to maximize the accuracy, while performing an exact computation of the $C_\ell$'s (no Limber approximation). Moreover in order to freely modify the used radial window function and verify the condition $W(z) \rightarrow W^2(z)$, we use the input option \texttt{userfile}. The other computational parameters are left as default.
On the \texttt{PySSC} setting side, we compute the wave mode integration using the precision parameter \texttt{precision} $=12$.

\begin{figure}
\centering
\includegraphics[width=.85\linewidth]{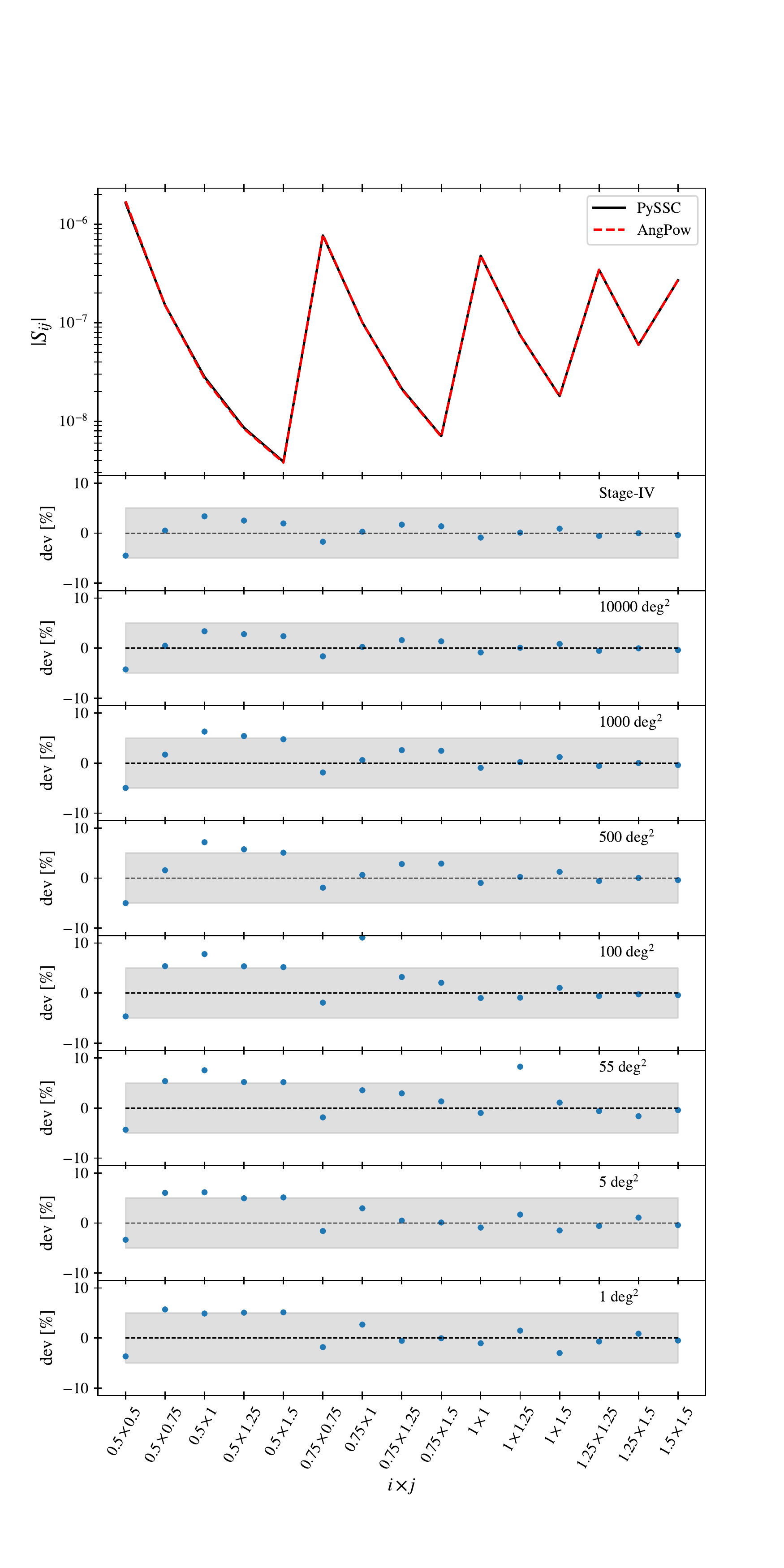}
\caption{Top panel : Absolute value of the $S_{ij}$ matrix elements in the case of a stage-IV angular mask, as predicted by \texttt{PySSC} in black solid line and \texttt{AngPow} in red dashed lines. The elements are ordered column by column of the lower half of the matrices, including the diagonal ($i \times i$). The corresponding x-coordinates $i\times j$ label which redshift bins are cross-correlated. Panels 2 to 9: relative deviations in percent between the two methods for decreasing survey area. The grey area depicts the 5\% discrepancy.}
\label{Fig:validation}
\end{figure}

In Fig. \ref{Fig:validation}, we compare the different $S_{ij}$ output matrices predicted by the two methods in the context of the eight angular masks previously introduced.
It shows a compatibility better than $\sim 6\%$ accuracy between the two $C_\ell$ computational methods for all masks larger than $1000$ deg$^2$ ($>2.5\%$ of the sky), while keeping a better than $\sim 10$\% accuracy for smaller patches.
Note finally that for each tested geometry, the deviation is damped for increasing redshifts.

\section{Galaxy distribution parameters}
\label{Appendix_nz}

In this appendix we show the value of the parameters of Eq.~\eqref{eq:nz1}, \eqref{eq:nz2} and \eqref{eq:nz3} we used to compute the three $n(z)$ considered in Sect.~\ref{Sect:surveynz}.

\begin{table}[!htb]
    \centering
    \caption{Value of the parameters describing the photometric redshift distributions.}
		\begin{tabular}{l | c c c c c c c }
			\hline
            {$n(z)$} & $c_b$ & $z_b$ & $\sigma_b$ & $c_0$ & $z_0$ & $\sigma_0$ & $f_{out}$\\
			\hline
			{Wide, $f_{out} = 0.1$} & $ 1.0 $ & $0.0$ & $0.05$ & $1.0$ & $0.1$ & $0.05$ & $0.1$\\
			\hline
			{Wide, $f_{out} = 0.25$} & $ 1.0 $ & $0.0$ & $0.05$ & $1.0$ & $0.1$ & $0.05$ & $0.25$\\
			\hline
			{Tight, $f_{out} = 0.1$} & $ 1.0 $ & $0.0$ & $0.02$ & $1.0$ & $0.1$ & $0.02$ & $0.1$\\
			\hline
		\end{tabular}
		\label{Tab:nz_app}
    \end{table}

\end{document}